\documentclass{article}

\usepackage[margin=10pt,font=small,labelfont=bf, labelsep=endash]{caption}
\usepackage[ruled,boxed]{algorithm2e}
\usepackage{amsmath}			

\SetAlFnt{\small}
\SetAlCapFnt{\small}
\SetAlCapNameFnt{\small}
\SetAlCapHSkip{0pt}
\IncMargin{-\parindent}

\newcommand \ra {\rightarrow}
\newcommand \oo {\infty}

\def \fns {\footnotesize}

\begin{document}

\centerline{\bf Efficient long division via Montgomery multiply}

\bigskip

\centerline{Last modified: 20 Aug 2016}
\bigskip
\centerline{Ernst~W.~Mayer}
\centerline{\fns 10190 Parkwood Dr.~\#1, Cupertino, CA 95014}
\centerline{\fns {\em ewmayer@aol.com}}

\bigskip

\begin{abstract}
We present a novel right-to-left long division algorithm based on the Montgomery modular multiply, consisting of separate highly efficient loops with simple carry structure for computing first the remainder $(x\ {\rm mod}\ q)$ and then the quotient $\lfloor x/q \rfloor$. These loops are ideally suited for the case where $x$ occupies many more machine words than the divide modulus $q$, and are strictly linear time in the ``bitsize ratio" $\lg x /\lg q$. For the paradigmatic performance test of multiword dividend and single 64-bit-word divisor, exploitation of the inherent data-parallelism of the algorithm effectively mitigates the long latency of hardware integer MUL operations, as a result of which we are able to achieve respective costs for remainder-only and full-DIV (remainder and quotient) of 6 and 12.5 cycles per dividend word on the Intel Core 2 implementation of the x86\_64 architecture, in single-threaded execution mode.  We further describe a simple ``bit-doubling modular inversion" scheme, which allows the entire iterative computation of the mod-inverse required by the Montgomery multiply at arbitrarily large precision to be performed with cost less than that of a single Newtonian iteration performed at the full precision of the final result. We also show how the Montgomery-multiply-based powering can be efficiently used in Mersenne and Fermat-number trial factorization via direct computation of a modular inverse power of 2, without any need for explicit radix-mod scalings.
\end{abstract}




\section{Introduction}
\label{sect:intro}

In an ingenious 1985 work \cite{Mont85}, Montgomery presented an algorithm for division-free modular multiplication (``modmul") which is ideally suited for modern binary compute hardware. For general background we refer the reader to Crandall \& Pomerance \cite{CP05}, who note that the method is a generalization of an old method of Hensel \cite{Hensel} for computing the inverses of 2-adic numbers, and who further give a nice survey of subsequent developments and algorithmic refinements, most of which are also detailed in the more-encyclopedic reference by Menezes {\em et al.} \cite{Mene96}. C\&P make specific note of the ``tough case" when the number $x$ is much larger than the modulus $q$ in the log-ratio sense, that is if $n$ is the smallest integer such that $x < q^n$, then we have $n \gg 1$. They remark on several highly-tuned computer algorithms for this case, all based on variants of the Barrett mod technique \cite{Bar87}. In the present work we show how the Montgomery modmul can be successfully brought to bear on both the general but especially the latter ($n$ large) case. The resulting division method consists of the following three steps, where $R$ is the binary arithmetic radix used for the core Montgomery modmul operations:
\begin{enumerate}
\item 	(Algorithms~A and B) Computes a scaled remainder $s$, related to the true remainder $r \equiv x$ (mod $q$) by $r = -s \cdot R^n$ (mod $q$) for Algorithm~A, and $r = +s \cdot R^{n-1}$ for Algorithm~B;

\item	(Algorithm~C) Efficiently computes the needed modular power of the radix $R$ via a divide-and-conquer scheme, and returns the true remainder $r$;

\item	(Algorithm~D) Takes $x$ and $r$ and produces the quotient $y = \lfloor x/q \rfloor$.
\end{enumerate}
Steps 1 and 3 are performed using simple loops which run through the elements of the (presumably multiword) vector $x$ in ascending order, and possess quite-similar loop-body arithmetic structure. Step 1 is all that is needed if a simple binary ``does $q$ divide x?" result is required. Step 2 uses similar core modmul arithmetic to compute the scaling factor needed to transform the Step 1 result out of ``Montgomery space", yielding the remainder $(x\ {\rm mod}\ q)$, which is used (in addition to $x$) as an input to Step 3. Steps 1 and 3 each require $O(n)$ size-$q$ modmul operations (one per loop execution), whereas Step 2 needs just $O(\lg N)$ modmuls. Thus, the scheme possesses the property that the work required to compute just the remainder $(x\ {\rm mod}\ q)$ is asymptotically half that needed to obtain both the remainder and quotient.

Let us first introduce some basic arithmetic and computational notation related to the various core multiply operations. Given nonnegative integers $x$ and $y$ whose product is desired with respect to some positive integer modulus $q$ (such that $0 \le x,y < q$), the Montgomery method begins with an arithmetic radix (or ``base") $R$, typically chosen for computational convenience. $R$ and $q$ must be coprime, thus the modular inverse $qinv$ of $q$ exists, such that
\begin{equation}
	q \cdot qinv \equiv 1\ ({\rm mod}\ R).
\end{equation}
The Montgomery multiply does not directly return $x \cdot y$ (mod $q$); rather it returns a ``shifted mod" involving the inverse of the radix $R$, thus we include $R$ in the argument list as a parameter:
\begin{equation}
	M(x,y,q;R) = x \cdot y \cdot R^{-1}\ ({\rm mod}\ q)
\end{equation}
which needs suitable adjustment if the ``true mod" $x \cdot y$ (mod $q$) is required. Often it is not~-- for example in the fast divisibility algorithm we shall give in the next section, we only care if the output is zero or not, obviating any need for postprocessing arithmetic~-- but when it is, the extra effort needed to scale the output in order to produce the true mod may be appreciable. The need to perform this scaling-away of the inverse power of the arithmetic radix is the major reason why the Montgomery multiply has not driven other modmul algorithms from the field in more or less any context requiring a modmul, although it has become ubiquitous in areas such as Diffie-Hellman and RSA-style public-key cryptography and elliptic curve arithmetic, all of which feature repeated multiplications with respect to a fixed modulus and thus in which radix-scalings are few relative to overall modmul count. In this respect long division $x/q$, where we especially mean ``long" in the literal sense that $\lg x/\lg q \gg 1$, appears to be a bit of an anomaly, because although it does feature repeated multiply modulo $q$, the best-known high-performance implementations such as that of the Gnu MP library \cite{GMP} eschew the Montgomery method, at least in any fashion resembling the one described here.

Here we take $R = 2^b$, a power of 2 typically equal to the an unsigned-integer wordsize $W$ supported by the compute hardware and programming language being used, or some integer power of $W$ if the modulus $q > W$, requiring a multiword-arithmetic implementation. It is further assumed that twos-complement integer arithmetic is being performed both with respect to $R$ and $W$. The restriction of coprimality of the radix and modulus thus simply reduces to that of the modulus $q$ being odd for the Montgomery modmul to work. The more general case of even moduli can easily be accommodated with simple pre- and postprocessing to take care of any powers of 2 appearing in $q$, which we shall also detail. (We specifically refer the reader to the end of the section describing Algorithm~D).

Assuming one has computed the needed mod-inverse $qinv$, a typical computational implementation of the Montgomery modmul consists of a three-multiply sequence, followed by a mod-subtraction, which can be expressed in pseudocode as\\
\\
{\bf Algorithm:} ${\rm MONT\_MUL_b}$\\
\IncMargin{5em} 
\begin{algorithm}[H]
\SetAlgoLined
\KwData{Unsigned b-bit integers $x,y,q,qinv$, $q$ odd and $q \cdot qinv \equiv 1\ ({\rm mod}\ 2^b)$}
\KwResult{The radix $R = 2^b$ Montgomery product $x \cdot y \cdot R^{-1}$ (mod $q$)}
\vspace{0.1in}
{
	$uint_b\ lo,hi$;\\
	$hi:lo = {\rm UMUL\_LOHI_b}(x, y)$;\\
	$lo = {\rm MULL_b}(qinv, lo)$;\\
	$lo = {\rm UMULH_b}(   q, lo)$;\\
	\eIf {$hi < lo$} {
		\Return $hi - lo + q$\;
	}{
		\Return $hi - lo$\;
	}
}
\label{MM}
\end{algorithm}
\vspace{0.1in}
%
%
%
Here $uint_b$ denotes a $b$-bit unsigned integer type (hardware-supported or emulated-multiword), and for the ${\rm UMUL\_LOHI_b}$ output pair we mimic the colon-separated register-pair notation used to describe the MUL instruction output in the x86\_64 instruction set architecture (ISA); arithmetically we mean $hi:lo = lo + R \cdot hi$.

\vspace{0.1in}
\noindent The three multiply variants appearing in the function are as follows:
\begin{table}[ht]
\begin{center}
\begin{tabular}{l p{3in}}
${\rm UMUL\_LOHI_b}(uint_b\ x, uint_b\ y)$	&	Full $2b$-bit double-width unsigned multiply which returns the lower and upper halves of $x \cdot y$ in a pair of variables. For this operation the result depends on whether the inputs are treated as signed or unsigned, so we add a leading `U' to specify that we mean the unsigned version.\\
\\
${\rm MULL_b}(uint_b\ x, uint_b\ y)$&	Lower-half multiply, that is, $x \cdot y\ ({\rm mod}\ R)$. We label the inputs as unsigned here, but for this operation the result is the same whether the inputs are treated as signed or unsigned.\\
\\
${\rm UMULH_b}(uint_b\ x, uint_b\ y)$&	Upper-half multiply, returns $\lfloor (x \cdot y)/R \rfloor$, the upper $b$-bit half of the unsigned product $x \cdot y$.
\end{tabular}
\end{center}
\end{table}
\setcounter{table}{0}

\noindent These three multiply operations are related simply as
$$
{\rm UMUL\_LOHI_b}(x,y) = x \cdot y = {\rm MULL_b}(x,y) + R \cdot {\rm UMULH_b}(x,y) .
$$
Since $R$ is a power of 2 we can also express MULL and UMULH in terms of simple binary shifts and masks of a full product:
\begin{align}
\nonumber	{\rm MULL_b}(x,y) &= {\rm UMUL\_LOHI_b}(x,y)\ \&\ (2^b - 1) ;\\
\nonumber	{\rm UMULH_b}(x,y) &= {\rm UMUL\_LOHI_b}(x,y) \gg b .
\end{align}
The three-multiply sequence ${\rm UMUL\_LOHI~/ MULL~/ UMULH}$ is followed by mod-$q$ subtraction of a pair of $b$-bit quantities, typically effected an if/else branch (as in our pseudocode), or uaing a conditional-mov-style operation, either of the explicit kind if the hardware supports it, or of the ``hand-rolled" bitwise-mask-and-add variety.


\section{Notes on the Newtonian Iterative mod-Inverse}
\label{sect:modinv}

The multiplication algorithm first requires computation of $qinv$, the multiplicative inverse of $q$ modulo $R$. This is a modular analog of the inverse over the reals, and in practice one even uses the same Newtonian iterative-inversion algorithm to compute $qinv$ as is commonly used to effect a fast floating-point inverse computation needing no hardware division, the latter being generally a slow operation in both the floating-point and integer domains. One starts with an initial guess $qinv_0$ having one or more low-order ``good bits" in the sense that $qinv_0 \cdot q \equiv 1\ ({\rm mod}\ 2^m)$ with $m \ge 1$. One advantage of the Newtonian iteration in the modular realm is that as long as even the least-significant bit of $qinv_0$ is correct, convergence is guaranteed. So since $q$ odd we can take $qinv_0$ to be any odd number, e.g.~1 or even $q$ itself. Montgomery (personal communication) alternatively suggests a kind of low-precision bitwise table-lookup in the form of a simple inline formula, yielding at least 5 correct low bits in the resulting initial iterate:
\begin{equation}
	qinv_0 = {\rm XOR}(3q, 2) .
\label{qinv0}
\end{equation}
In a 32 or 64-bit context one can save a further few cycles via use of a precomputed 128-element bytewise lookup table indexed by $(q\ \&\ {\tt0xFF})\gg1$. The ensuing Newton iteration involves repeated steps of form
\begin{equation}
	qinv_{j} = qinv_{j-1}*(2 - q*qinv_{j-1}),
\label{newt}
\end{equation}
Where the asterisk operator is a shorthand for the ${\rm MULL_b}$ operation (that is, multiplication modulo $2^b$) on the pairs of operands it separates, either with respect to some fixed arithmetic base $2^b > q$, or to some variable power of 2 whose number of bits increases with each iteration to keep pace with the per-iteration doubling of number of converged bits of $qinv$, and which exceeds $q$ only on the final iteration. The advantage of taking $R$ to be a power of the natural integer wordsize is apparent here, since e.g.~if $R = W$ we really can use the hardware multiply operation denoted in programming languages such as C and Fortran by `$*$', that is, integer product modulo $W$. For $R > W$ we replace the literal `$*$' by a software-emulated ${\rm MULL_b}$ operation, which for $R$ not terribly larger than $W$ (say, $R < W^{20}$) will typically be based on some hardware-optimized implementation of grammar-school multiply. For larger radices, optimized software implementations will transition to a subquadratic algorithm such as Toom-Cook and eventually for very large operands, to some kind of fast transform-based discrete convolution. Similar large-argument optimizations apply to the other forms of multiply required by the Montgomery algorithm.

Returning to the inverse computation, the number of significant bits at the bottom doubles on each iteration, until we reach the bitwidth set by the modulus $q$, which determines the width of the ${\rm MULL_b}$ operation used. Since we only care about the final converged value of $qinv$ we can do the iteration ``in-place". A typical C implementation of the inversion~(\ref{newt}) for $R = 2^{64}$ might look as follows, where $uint_{64}$ is a typedef for the native unsigned 64-bit integer type on the platform in question:\\
\\
{\bf Algorithm:} {64-bit mod-inverse computation}\\
\begin{algorithm}[H]
\KwData{Unsigned 64-bit odd integer modulus $q$}
\KwResult{Mod-inverse $qinv$, such that $q \cdot qinv \equiv 1\ ({\rm mod}\ R = 2^{64})$}
\vspace{0.1in}
	{
	$uint_{64}\ tmp,\ qinv = {\rm XOR}(3q, 2)$;\\
	\For {$j\leftarrow 1$ \KwTo $4$}{
		$tmp  = {\rm MULL_{64}}(q, qinv)$;\\
		$qinv = {\rm MULL_{64}}(qinv, (uint_{64})2 - tmp)$;\\
	}
}
\end{algorithm}
\vspace{0.2in}
%
%
%
Let us crunch the numbers on an actual numerical example, the 64-bit modulus $q = 16357897499336320049$. First note that $3q \equiv 19$ (mod 16), which in binary $= 10011_2$. XOR of this with 2 toggles the set bit in the 2s slot to yield $10001_2$. Since we only care about these 5 bits of the initial inverse seed value, we could simply initialize $qinv_0 = 17$. In fact, since in this case we are targeting a 64-bit result the high bit of this initial iterate gains us nothing in terms of work savings: We could equally well take $qinv_0 = 0001_2 = 1$, and in either case would still require 4 iterations to obtain a 64-bit-converged result. However, since in practice such bit-masking is superfluous, we illustrate the algorithm exactly as given. We thus have XOR(3q, 2) = 12180204350589856915 (note the multiply by 3 here is also modulo $2^{64}$) which in hexadecimal = A908C752C8936C9\underline{1}, where we have underlined the converged hex digits, which at this point means just the rightmost one (though at least the low bit of the next-higher hex digit is also converged). We note that for this particular choice of $q$ our initial-iterate formula actually yields 6 good bits in $qinv_0$. (In general there is a 50\% chance of this ``extra bits for free" behavior occurring, according to the low bits of $q$). We now iterate 4 times, obtaining the data in Table~\ref{table_iter64}, again with the same underline-highlighting of converged digits. The right-column iterate $qinv_j$ is the input (in addition to the fixed $q$) to the next iteration. The iteration yields the desired 64-bit mod inverse, which in decimal form is $qinv = 9366409592816252113$. Were we to do one more iteration, we would observe that the product $q \cdot qinv$ would have its 64-bit lower half equal to unity, confirming that $qinv$ has reached its desired fixed point.

\begin{table}[ht]
\begin{center}
\caption{\label{table_iter64}Convergence history of Newtonian iteration for 64-bit mod-inverse of $q = 16357897499336320049$:}
\begin{tabular}{c|c|c|c}
$j$	&	$q \cdot qinv_{j-1}$	&	$qinv_j$ &	$bits$\\
\hline
$1$	& {\tt F3D66805FF3D2BC\underline{1}}	& {\tt 4141B6714938A\underline{4D1}}	&	$12$	\\
$2$	& {\tt B0955EB99F05F\underline{001}}	& {\tt 656E0C4A27\underline{9FB4D1}}	&	$24$	\\
$3$	& {\tt 0A62BFBCBF\underline{000001}}	& {\tt 6EAB\underline{2BE6389FB4D1}}	&	$48$	\\
$4$	& {\tt E97F\underline{000000000001}}	& {\tt \underline{81FC2BE6389FB4D1}}	&	$64$
\end{tabular}
\end{center}
\end{table}
%
%
%
%

\vspace{-0.1in}
An alternative to the XOR-based initialization is to instead use a small precomputed bytewise lookup table yielding 8 bits for the initializer, followed by just 3 iterations to yield 64 bits. 
If the hardware has fast hardware support for 32-bit integers (as is the case on many platforms still running 32-bit operating systems), it makes sense to begin using 32-bit operands, followed by a single 64-bit iteration. Similarly, if a radix larger than 64 bits is used, one can proceed to 64 bits in $qinv$ using the above loop, then do a single iteration at 128 bits using software-emulated 128-bit integer arithmetic, and so forth. As the operands get large enough to exceed the natural hardware wordsize $W$ it also pays to exploit some simple optimizations resulting from the properties of the mod-inverse. For example, let us say our hardware supports integers of 64 bits, in the sense that the operations ${\rm MULL_{64}}$ and ${\rm UMULH_{64}}$ (either directly or in emulated form via the upper half of ${\rm UMUL\_LOHI_{64}}$) are supported by way of hardware instructions. Further assume we have a 64-bit-accurate mod-inverse in hand (that is, bits 0:63 of the inverse), and wish to do one more Newtonian modular inversion iteration modulo $2^{128}$ to obtain a 128-bit mod-inverse. We can do so efficiently, by way of direct computation of bits 64:127 of the inverse, using a combination of three of the above 64-bit ${\rm MUL}$ primitives. More generally, given a $b$-bit-converged mod-inverse, one can obtain $b$ more bits of precision using just three $b$-bit ${\rm MUL}$ primitives. Here the bit count $b$ need not be equal to that of the ensuing Montgomery-mul arithmetic~-- all the analysis holds generally~-- but let us assume it for notational ease.

Define an emulated $2b$-bit type (e.g.~in C via a struct) consisting of a pair of $b$-bit ``digits" $d_0$ and $d_1$, such that a $2b$-bit quantity $x$ expands~-- here again using C-style notation~-- as $x = (x.d_1 \cdot R + x.d_0)$, where $R = 2^b$. We then take our $b$-bit-converged mod-inverse $qinv_b$ and use it to initialize the low word of such a $2b$-bit variable, that is, set $qinv.d_0 = qinv_b$ and $qinv.d_1 = 0$. Since we presumably are working with a $2b$-bit modulus q, we similarly store it in another $2b$-bit variable $q = (q.d_1 \cdot R + q.d_0)$~-- the $b$-bit inversion will have only used the lower half of this. We also have by construction that the fully (that is, $2b$-bit) converged inverse satisfies ${\rm MULL_{2b}}(q, qinv) \equiv 1\ ({\rm mod}\ R)$. We can take advantage of these two facts to speed the $2b$-bit inversion iteration. The product $q\cdot qinv_b$ yields a $2b$-bit partial multiplicative identity, where by ``partial" we mean having only its lower $b$ bits unity, that is, identity modulo $R$:
$$
	I_b := q \cdot qinv_b = (q.d_1 \cdot R + q.d_0) \cdot qinv.d_0 = q.d_1 \cdot qinv.d_0 \cdot R + q.d_0 \cdot qinv.d_0 \equiv 1\ ({\rm mod}\ R),
$$
where here the `$\cdot$' are shorthand for full-width products of the operands in question. In terms of $b$-bit hardware operations, since our input value $qinv_b$ is $b$-bits-good, we know {\em a priori} that the lower half (bits $0:b-1$, word $d_0$) of the $2b$-bit product $I_b$ is equal to 1, and need not explicitly compute this half. Extracting bits $b:2b-1$ of $I_b$, which is the half of interest, we have:
$$
	I_b.d_1 = {\rm MULL_b}(q.d_1, qinv.d_0) + {\rm MULH_b}(q.d_0, qinv.d_0),
$$
where the addition is simple $b$-bit hardware integer addition modulo $R$. To effect the subtraction of the 2-word result $I_b = (1, I_b.d_1)$ from the constant 2 we note that this subtraction leaves the low $b$-bit word unchanged so long as $b \ge 2$, i.e. we have at least a 2-bit-converged mod-inverse. The high word simply requires arithmetic negation, thus $(2-I_b) = (1, -I_b.d_1)$. We use that the lower half of this is unity to simplify the ensuing ${\rm MULL_{2b}}(qinv, (2-I_b))$ operation: since $qinv$ still just has nonzero low word, one has
$$
	qinv \cdot (2-I) = qinv.d_0 \cdot (-I.d_1 \cdot R + 1).
$$
The low half of this is again guaranteed to be unity, so starting with a $b$-bit accurate approximation to the multiplicative mod-inverse stored in $qinv.d_0$, the entire sequence of operations needed to perform the next Newton iteration and obtain a $2b$-bit result can be condensed to the following direct computation of the next-higher $b$ bits of the inverse, needing just three $b$-bit multiply instructions, along with one addition and one arithmetic negation:
\begin{equation}
	qinv.d_1 = {\rm MULL_b}(-qinv.d_0, {\rm MULL_b}(q.d_1, qinv.d_0) + {\rm UMULH_b}(q.d_0, qinv.d_0)) .
\label{next_b}
\end{equation}
where the negation and addition are both twos-complement, i.e. modulo $R$.

In order to compute a work estimate for~(\ref{next_b}), we must address the question of whether the upper-half product here is a true half-multiply (in the sense of needing asymptotically the same computational work as the a MULL operations as $b$ becomes large) or whether the UMULH needs to be computed using a full double-width UMUL\_LOHI (retaining just the upper half of the result) in order to properly capture the carry out of the low half of the product. In general, half-multiply implementations of UMULH for multiword inputs proceed by truncating the multiplication rhombus and approximating the exact carry out of the the lower-half product using a ``carry layer" of thinness similar or equal to the bitwidth of the hardware wordsize $\lg W$. This is effective if the carries occurring in the wordsize computations are local, which is a good assumption if the inputs are quasirandom, in which case the chance of an incorrect carry resulting from the truncated-rhombus approximation is in some rough sense proportional to $1/W$. In the context of the Newton iteration for the mod-inverse, however, the inputs to the $q \cdot qinv$ product yielding our partial multiplicative identity are the very opposite of random in this sense, because the very purpose of the iteration is to produce a product whose low $b$ bits have the specific pattern 000...001. This by definition entails ``maximally nonlocal" carry behavior in the low half of the $q \cdot qinv$ product. But that very same property provides a way out of the seeming difficulty, since it guarantees that the discarded low-half product ${\rm MULL_b}(q.d_0, qinv.d_0) = 1$, by construction. In other words, the UMULH in the streamlined iteration~(\ref{next_b}) can truly be performed as a half-multiply, but if one does things this way one must compare the terms summed in the carry-layer approximation (which sum is normally simply discarded after it has been used to generate the carry) to the expected value 0, in order to identify any missed carry resulting from the approximation. We shall enounter a similar application of error-corrected upper-half multiply in our remainder and quotient algorithms.

Thus the cost of~(\ref{next_b}) can be brought down to three $b$-bit half-multiply instructions, which is less than one-half the MUL cost needed for the na\"ive $2b$-bit version of the inversion iteration,\ for which the inner $I = {\rm MULL_{2b}}(q,qinv)$ operation forms a $2b$-bit result using one double-width ${\rm UMUL\_LOHI_b}$ operation and two $b$-bit ``half-multiply" (with respect to the width of the $2b$-bit inverse we seek) ${\rm MULL_b}$ instructions:
\begin{align}
\nonumber	&I.d_0:d_1 = {\rm UMUL\_LOHI_b}(q.d_0, qinv.d_0) ;\\
\nonumber	&I.d_1 +\!\!= {\rm MULL_b}(q.d_0, qinv.d_1) + {\rm MULL_b}(q.d_1, qinv.d_0) .
\end{align}

This then has its high digit negated to produce $(2 - I)$ prior to being fed to a second such multiply of the ${\rm MULL_{2b}}(I,qinv)$. Analogous considerations apply to each additional bit-doubling iteration required, so the work of each successive ``bit-doubling" step using~(\ref{next_b}) is asymptotically equivalent to the summed work of all the preceding iterations, hence the total work needed to obtain the fully converged inverse using this method is indeed less than that needed by a single ``na\"ive" Newtonian iteration performed at the full final precision needed.

By way of example, let us compute the inverse of the 128-bit $q = 225797717267637708506527464987314161.$  Using 64-bit hardware arithmetic we have $q.d_0 = 1654746039858251761$ and $q.d_1 = 12240518780192025$. We compute $qinv$ with respect to the arithmetic modulus $2^{128}$, the smallest power of the hardware wordsize greater than $q$. Starting with the initial guess formula~(\ref{qinv0}), 4 Newton iterations yield a 64-bits-good partial inverse $qinv.d_0 = 18061898331188349201$, and it is easily verified that ${\rm MULL_{64}}(q.d_0, qinv.d_0) = 1$. Now making use of~(\ref{next_b}) with operand bitsize $b = 64$:
\begin{align}
\nonumber	&qinv.d_1 = {\rm MULL_{64}}(-qinv.d_0, {\rm MULL_{64}}(q.d_1, qinv.d_0) + {\rm UMULH_{64}}(q.d_0, qinv.d_0))	\\
\nonumber	&\qquad = {\rm MULL_{64}}(-qinv.d_0, 12364022002462652329 + 1620223851777327935)		\\
\nonumber	&= {\rm MULL_{64}}(384845742521202415, 13984245854239980264) = 5329826773734796952 .
\end{align}
Thus $qinv\!=\!5329826773734796952\!\cdot\!2^{64}\!+\!18061898331188349201$, and again ${\rm MULL_{128}}(q, qinv)\!=\!1$, as desired.


\section{A Fast Multiword Divisibility Test}
\label{sect:algo_a}

In this case it is not modmul by another number less than the radix we are interested in, but rather modmul by the radix $R$ itself. In a typical modular reduction of a multiword integer of form
\begin{equation}
	x = \sum_{j=0}^{n-1} a_j R^j,
\label{exp_x}
\end{equation}
one might start with the most-significant digit $a_{n-1}$, reduce $a_{n-1} \cdot R$ modulo $q$ and carry the result rightward into the next-lower digit, a ``left-to-right" approach to the reduction if one visualizes the terms of the above expansion written out in decreasing significance from left to right, analogously to the way one writes multi-digit numbers in decimal and other common small bases.

Here we exploit the fact that the Montgomery modmul naturally introduces an inverse power of the radix $R$ every time it is performed. Thus, starting with the least-significant digit $a_0$, a Montgomery modmul of this with unity yields $M(a_0,1) \equiv a_0 \cdot R^{-1}\ ({\rm mod}\ R)$. But this is exactly the ``modular reduction carry" one needs if proceeding in right-to-left fashion, that is, the carry into the next-higher term. That observation, followed by a bit of carry-related work, leads immediately to the fast multiword divisibility algorithm captured in pseudocode form as Algorithm~A.

\begin{algorithm}
\SetAlgoLined
\SetAlgoRefName{A}
\KwData{Unsigned b-bit integer array $x[n]$, b-bit scalars $q,qinv$, with $q$ odd and $q \cdot qinv \equiv 1\ ({\rm mod}\ 2^b)$}
\KwResult{True if $q$ divides $x$, false otherwise.}
\vspace{0.1in}
{
	$uint_b\ tmp,\ bw,\ cy = 0$;\\
	\For {$i\leftarrow 0$ \KwTo $n-1$}{
		$tmp = x_i - cy$;\\
		$bw = (cy > x_i)$;\\
		{\em// Add q to loop output if had a borrow:}\\
		$tmp = {\rm MULL_b}(tmp, qinv) + bw$;\\
		$cy = {\rm UMULH_b}(tmp, q)$;\\
	}
	\Return $(cy == 0)$\;
}
\caption{IS\_DIV\_A, fast right-to-left divisibility check}
\label{algo_a}
\end{algorithm}

%
%
This algorithm can be trivially generalized to also permit even divisors, by first counting trailing binary zeros in $q$ (call this number $tz_q$), and one immediately returns FALSE if $x$ has fewer trailing zeros than $q$. Otherwise~-- in pseudocode but with all quantities arbitrarily large here~-- one returns IS\_DIV\_A($x',\ q',\ qinv,\ n')$, where  $x' = (x \gg tz_q)$, $q' = (q \gg tz_q)$ and $qinv$ is the mod-$R$ inverse of the right-justified divisor $q'$. The number-of-words parameter can be set via $n' = n - (tz_b \gg b)$ if one wants to account for the length reduction due to the right-shifting of inputs by one or more entire $b$-bit words, without the extra expense of counting leading zeros of $x$ required if one wants to achieve the maximum possible length reduction by counting partial-word shifts; otherwise one can simple take $n' = n$.

We assume nothing about hardware support for carry flags on add and subtract here, thus the explicit check for borrow-on-subtract using unsigned comparison $(cy > x_i)$. The one slight subtlety related to the case where the subtraction results in a borrow is that since the underlying remainder accumulation is $({\rm mod}\ q)$, such a borrow must be accounted for by re-adding $q$. Since the ensuing low-half multiply of $tmp$ is by $qinv$ and $q*qinv \equiv 1\ ({\rm mod}\ R)$, one can simply add 1 to the ${\rm MULL_b}(tmp, qinv)$ result instead of adding $q$ to the computed difference stored in $tmp$, which is advantageous if $q$ occupies multiple machine words.

We can ignore the possible carry out of the resulting $tmp = {\rm MULL_b}(tmp, qinv) + bw$ step, because we are computing a result $({\rm mod}\ q)$: If the addition overflows, $tmp$ ends up being $R = 2^b$ too small as a result of twos-complement arithmetic. If we restore the dropped $R$ in the next step, instead of
$$
cy = {\rm UMULH_b}(tmp, q)
$$
we have
$$
cy = {\rm UMULH_b}(tmp + R, q) = {\rm UMULH_b}(tmp, q) + q ,
$$
i.e. the result is the same $({\rm mod}\ q)$. Thus we are letting the twos-complement arithmetic help with the modding here. (Note that the ${\rm MULL_b}$ operation is doing effectively the same thing).

Because of the multiply-by-unity nature of the processing of each term in the expansion of $x$ we do not need the double-width multiply which opens the three-multiply sequence of the general Montgomery modmul: Here (ignoring the carry-in from the next-lower term for the moment) we have $hi:lo = {\rm UMUL\_LOHI}(x_i, 1) = 0:x_i$, and instead of explicitly negating the output of the UMULH operation to effect the final $hi - {\rm UMULH}()$ step, we simply subtract the UMULH result from the next-higher term at the start of the next loop execution, thus inlining the negation with the leftward carry propagation. The core loop in the algorithm thus constructs a scaled sequence of modular partial sums (negated Hensel remainders) such that on the $i$th loop iteration,
\begin{equation}
	cy = - R^{-(i+1)} \sum_{j=0}^{i} a_j R^j,\ {\rm for}\ i = 0, ... , n-1.
\label{Asum}
\end{equation}
To obtain the true remainder $r := (x\ {\rm mod}\ q)$, we must take the output value, scale it and negate it $({\rm mod}\ q)$:
\begin{equation}
	r = -cy \cdot R^n \ ({\rm mod}\ q).
\label{Ascale}
\end{equation}
Thus if we desire the true remainder rather than merely knowing whether it is zero on not, we just encapsulate the same computational loop in a slightly different interface, which here further performs the negation prior to returning to reinforce that it is a $({\rm mod}\ q)$ negation rather than a twos-complement $({\rm mod}\ R)$ one:
\vspace{0.1in}

\begin{algorithm}[H]
\SetAlgoLined
\SetAlgoRefName{A'}
\KwData{As for Algorithm A}
\KwResult{Scaled remainder $s \equiv x \cdot R^{-n}$ (mod $q$).}
\vspace{0.1in}
{
	{\em [Same as loop body of Algorithm A]}\\
	\eIf {$cy \ne 0$} {
		\Return $q - cy$\;
	}{
		\Return $cy$\;
	}
}
\caption{REMAINDER\_A, fast right-to-left scaled remainder computation}
\label{algo_ar}
\end{algorithm}
\vspace{0.1in}

By way of a concrete numerical example which can be easily reproduced either using a tiny program or ``by hand" (using e.g.~the Unix `bc' utility or a freeware package such as Pari/GP), let us consider the problem of finding the remainder of $x = 2^{977}-1$ (which needs $n = 16$ 64-bit words to store explicitly) with respect to the 64-bit modulus $q = 16357897499336320049$. The 64-bit mod inverse $qinv = 9366409592816252113$, and Table~2 lists the successive UMULH outputs at end of each loop execution in the middle column, and the results of applying the above scaling formula (with loop iteration $i$ replacing $n$ in the radix power) to these successive iterates in the third column from left~-- we write the scaled outputs in hexadecimal form to ease comparison with analogous data we shall provide in \S~\ref{sect:algo_b}. The reader can verify that these right-column data equal $2^{64}-1, 2^{128}-1, 2^{192}-1, ...$ (mod $q$), i.e. that the iterates are indeed the scaled partial sums as given in~(\ref{Asum}). Note that if we desire the Hensel quotient, we simply save the MULL outputs within the loop, but must further account for the possible borrow from the subtraction; these modifications produce the same quotient loop as given in our true-quotient-producing Algorithm~D, but with no true-remainder subtraction step preceding the loop.

\begin{table}
\begin{center}
\label{table_a}
\caption{Per-loop-iteration data for Algorithm A applied to $x = 2^{977}-1$ and $q = 16357897499336320049$, with $R=2^{64}$. Rightmost column: A simple x86\_64 assembly-code implementation of the remaindering loop.}
\begin{tabular}{r|r|c| l l}
	$i$	&Result ($cy$)\qquad\qquad	&$-cy \cdot R^{i+1}$ (mod $q$)\qquad\\
\hline
	 0	& 8052108280172618802	&{\tt 1CFD12E467CEDBCE	}	&	\qquad{\tt	MOV	}	&	{\tt [x]\ \ \ ,R10}\\
	 1	&13395404783617144454	&{\tt 4D611EA3809531E7	}	&	\qquad{\tt	MOV	}	&	{\tt [len]\ ,RCX}\\
	 2	&14290423936650903017	&{\tt BD293142B725F2B8	}	&	\qquad{\tt	MOV	}	&	{\tt [q]\ \ \ ,RSI}\\
	 3	&12694450473754035419	&{\tt 6DA6C745723D2042	}	&	\qquad{\tt	MOV	}	&	{\tt [qinv],RBX}\\
	 4	&13022541340536637118	&{\tt 62A5DDA094A31338	}	&	\qquad{\tt	XOR	}	&	{\tt RDX,RDX	// cy = 0}\\
	 5	& 7849884873665013561	&{\tt 32D35CC447414DD0	}	&	\qquad{\tt	XOR	}	&	{\tt RDI,RDI	// bw = 0}\\
	 6	&  139461722106114244	&{\tt C4E358F892AFD7CB	}	&	{\tt loop:		}	&	{\tt // Loop downward}\\
	 7	& 6660703926365324543	&{\tt 841646B12435044D	}	&	\qquad{\tt	MOV	}	&	{\tt (R10),RAX	// x[i]}\\
	 8	&13147792529020392181	&{\tt 953C002AC7E9CA9B	}	&	\qquad{\tt	ADD	}	&	{\tt 8,R10	// \&x[i+1]}\\
	 9	&12940374201018017432	&{\tt 87A198DD565BAE6E	}	&	\qquad{\tt	SUB	}	&	{\tt RDX,RAX	// x[i]-cy}\\
	10	& 9345504322264630629	&{\tt 665B982F2C57CF9F	}	&	\qquad{\tt	SETC}	&	{\tt DIL	// bw = CF}\\
	11	&10439060481633841924	&{\tt C2F464196442F72D	}	&	\qquad{\tt	IMUL}	&	{\tt RBX,RAX 	// tmp*qinv}\\
	12	&11180607432989656657	&{\tt 92AB735A1C3B4927	}	&	\qquad{\tt	ADD	}	&	{\tt RDI,RAX	// tmp+=bw}\\
	13	& 6407570042918850368	&{\tt C2A0BA67701C6754	}	&	\qquad{\tt	MUL	}	&	{\tt RSI	// UMULH(q,tmp)}\\
	14	&12260751538328612790	&{\tt 339D02265F21100D	}	&	{\tt DEC RCX	}	&	{\tt }\\
	15	& 5031209829575536552	&{\tt 77ABEA1607BF1817	}	&	{\tt JNZ loop	}	&	{\tt // loop if RCX != 0}\\
		& 						& 							&	\qquad{\tt	MOV	}	&	{\tt RDX,[cy]	// Output}\\
\end{tabular}
\end{center}
\vspace{-0.4in}
\end{table}
The output of the loop is $cy = 5031209829575536552$, which is nonzero and thus indicates that $q$ does not divide $2^{977}-1$, and indeed the properly scaled output $8623243291871090711 \equiv 2^{977}-1$ (mod $q$).

Timing tests of Algorithm~A on a commodity 64-bit PC~-- the author used a 64-bit GCC 4.2 build on a vintage 2009 sub-\$1000 Apple Macbook with a 2 GHz Intel Core 2 processor~-- yield a timing of 17 cycles per loop iteration (that is, per dividend word) in the case of a 64-bit modulus. This is for a generic C implementation, whose only hardware-specific optimization was inlining of the x86\_64 MUL instruction (128-bit product of 64-bit unsigned integer inputs) via a tiny assembly code macro in order to provide the needed UMULH functionality. Obvious optimizations to consider with a view toward speeding things up further include explicit unrolling of the loop, a version of the function which tests several candidate moduli in parallel, and coding of either the entire loop body or function in assembler on one's target architecture. On x86\_64, this can further include usage of the hardware carry flag to eliminate the unsigned comparisons used to check for carry-on-add and borrow-on-subtract in our pseudocode examples. However, after trying all of these, the only one which yielded any appreciable throughput improvement was the multiple-modulus approach, which does not appear to be relevant to the ``bread and butter" case of dividing a large number $x$ by a single-word $q$, so-called ``$n/1$ division" in the parlance of~\cite{GMP_div}. Fortunately, first appearances oft deceive; we shall have more to say about this shortly.

In a very real sense a single-word-$q$ is the worst-case scenario for the algorithm, because the serialization imposed by the arithmetic structure imposes a severe constraint, and the loop body is so simple that there is little other work which can be performed while waiting for the two high-latency MUL operations to complete. This also explains why loop unrolling is not helpful in this case. Things should be better for multiword $q$, since the attendant multiplies can be structured to hide the latency of the underlying hardware MUL operations.

The rightmost column of Table~2 contains a simple x86\_64 assembly-code implementation of the remaindering loop, with operand pairs in SRC, DEST order; we further left-justify loop-control-related labels and instructions and tab-indent the others. It is not the assembly code itself which is of interest here but rather what it tells us about the expected cycle count per loop execution. The first 4 lines simply load the address of the $x$-array, the array length and the integers $q$ and $qinv$ into hardware registers, where we give full play to the historical arcana contained in the x86\_64 instruction set architecture; the next 2 lines zero a pair of carry-related registers. The key 7-instruction sequence is the one which follows the `{\tt loop:}' label and implements the loop body: (1) dereference array pointer to move current dividend word into register R10; (2) increment array pointer by 8 bytes to point to next word of $x$; (3) subtract lower-word carry from current dividend word (this sets the internal x86 carry flag bit CF if it underflows); (4) use the SETC instruction to save the state of the carry flag in DIL, the low byte of RDI (i.e. RDI contains $1$ if $x_i - cy$ underflowed, 0 if not); (5) low-half-multiply of $tmp$ and $qinv$; (6) add the saved borrow to the MULL result; (7) high-half-multiply. With regard to the latter instruction, note the x86 MUL instruction assumes one input is in register RAX and takes the register containing the other input as its only explicit argument; it then overwrites the RDX:RAX register pair with the 2-word result.

The key performance-limiting dependency chain here is that consisting of instructions SUB-IMUL-ADD-MUL~-- the carry-saving instruction, be it SETC as here or SBB as in our later quotient-loop example, can be run parallel to IMUL~-- which according to widely available Core 2 instruction-timing data~\cite{agner} have respective latencies which sum to 1+5+1+8 = 15 cycles, just 1-2 cycles less than  observed in the timing experiments on our builds, which use $x$ vectors several thousand words in length, sufficient to render any non-loop-related overhead negligible.

Our GCC-built high-level C-code and assembly versions of the loop yield respective timings of 17 and 16 cycles, closely matching the best-expected one on the Core 2 implementation of x86\_64, and the analysis reveals that there is no possible clever restructuring of the code which can avoid the issue of instruction latency. This also explains why the aforementioned multiple-modulus variant of the code yielded a gain: processing more than one modulus in parallel allows useful work to be done in the idle cycles of the single-$q$ instruction chain. The case of a fixed dividend and multiple divisors is common enough to be of some interest, but if possible we would like to be able to make use of this insight regarding overlapping of independent sequences in the single-$q$ case as well. For $x$-array dimensions of sufficient size~-- say 10 or more elements~-- we can effectively do so by breaking the single processing loop into 2 or more disjoint segments, each of which computes a partial remainder for just its portion of the summation~(\ref{exp_x}). These independent remaindering steps can be performed in parallel and their outputs combined at the end. By ``parallel" we mean not in the usual sense of being run on separate processors or in separate execution thread, but the interleaved execution is quite akin to multithreading, just in the sense that the ``added threads" are really idle cycles in an existing single-threaded execution stream. Because this loop restructuring is quite distinct from unrolling (although it can be combined with the latter process, if desired), we prefer the term ``loop folding", and define the number of array segments which get processed in parallel as the folding factor $F$. We shall also use the term ``$F$-way folding" to describe a particular value of $F$. For example, a 2-way-folded implementation might look like that shown in Algorithm~\ref{algo_ax2}. If the dividend-array dimension is not an exact multiple of $F$, we simply add one or more zero-padding elements beyond $x_{n-1}$ in order to make it so.

\begin{algorithm}
\SetAlgoRefName{Ax2}
\SetAlgoLined
\vspace{0.1in}
{
	$uint_b\ n2 = n/2,\ tmp0,\ tmp1,\ bw0,\ bw1,\ cy0 = 0,\ cy1 = 0$;\\
	\For {$i\leftarrow 0$ \KwTo $n2-1$}{
		$tmp0 = x_i - cy0;						\qquad\qquad\qquad\qquad\ \ \ 	tmp1 = x_{i+n2} - cy1;	$\\
		$bw0 = (cy0 > x_i);						\qquad\qquad\qquad\qquad\ \ 	bw1 = (cy1 > x_{i+n2});	$\\
		{\em// Add q to loop output if had a borrow:}\\
		$tmp0 = {\rm MULL_b}(tmp0, qinv) + bw0;\ \ tmp1 = {\rm MULL_b}(tmp1, qinv) + bw1;	$\\
		$cy0 = {\rm UMULH_b}(tmp0, q);			\qquad\qquad\ \ \ 	cy1 = {\rm UMULH_b}(tmp1, q);	$\\
	}
	$uint_b\ scale = ${\em [Compute scaling factor $P\cdot R \equiv R^{n2+1}$ (mod $q$); cf.~Algorithm C] }\\
	$cy1 \leftarrow {\rm MONT\_MUL_b}(cy1,\ scale,\ q,\ qinv)$;	{\em// $cy1 * P$ (mod $q$)}\\
	{\em // Sum the scaled partial remainders:}\\
	$uint_b\ cy = (cy0 + cy1)$;\\
	{\em // Check for overflow on addition:}\\
	\If {$(cy < cy0)\ or\ (cy \ge q)$} {
		$cy \leftarrow (cy - q)$;
	}
	\Return $(cy == 0)$\;
}
\caption{2-way loop-folded implementation of algorithm IS\_DIV\_A.}
\label{algo_ax2}
\end{algorithm}

\noindent After the loop completes, we combine the $F$ partial remainders:
\begin{equation}
cy = cy_0 + cy_1\cdot P + cy_2\cdot P^2 + ... + cy_{F-1}\cdot P^{(F-1)}\ \ ({\rm mod}\ q),
\label{algo_a_part}
\end{equation}
where the $x$-array is presumed to have been zero-padded to make $n$ a multiple of $F$ and the modular radix power $P = R^{n/F}\ ({\rm mod}\ q)$ can be efficiently computed using Algorithm~C, described in~\S~\ref{sect:algo_c}. The weighted summation can be efficiently performed by rearranging it using the standard nesting, whereby one computes the pairwise weighted-sums beginning with the innermost and working outward:
\begin{equation}
cy = cy_0 + (cy_1 + (cy_2 + ... + (cy_{F-2} + cy_{F-1}\cdot P)\cdot P)\ ...\ \cdot P)\ \ ({\rm mod}\ q).
\label{algo_a_nest}
\end{equation}
The result is then negated (mod $q$) and scaled using a final modmul by $P$ using the identity~(\ref{Ascale}) if the true remainder is desired. This procedure requires computation of just a single modular power of the radix $R$; note that if one also uses a Montgomery-mul to perform the scaling multiplies as shown in our 2-way pseudocode example, one computes not $P\ ({\rm mod}\ q)$ but rather $P\cdot R = R^{n/F+1}\ ({\rm mod}\ q)$.

In this context, the same instruction-sparseness which causes the left-to-right remaindering algorithm to be latency-dominated in serial-execution mode now proves to be an advantage, since it yields much opportunity for idle-cycle-filling interleaving.  We tried several folding factors, first using C code; again using the single-word-$q$ case on the Core 2, for $F$ = 1, 2 and 4 we obtain average cycle counts per $x$-word of 17, 9.5 and 6.5 cycles, respectively. We have already established that the strictly serial loop has a minimum expected cycle count of 15, so the for 2-way and 4-way folding the best-case cycle counts are 15/2 = 7.5 and 15/4 = 3.75. These~-- especially the last one~-- are almost certainly optimistic, as they disregard additional factors such as instruction-decode and issue-rate limits. In practice, our pure-assembler version of the respective loops yields a mere 1-cycle gain for $F=1$ and only modest gains of one-half-cycle per input word for the larger values of $F$, thus our current best timings are 16, 9 and 6 cycles per $x$-word. We did not attempt to go beyond $F = 4$ since at that value the interleaved-instruction code uses up nearly all of the available general-purpose hardware registers of the x86\_64 architecture. Larger folding factors are certainly possible but will require very careful analysis to reduce the need for register spills, register-freeing optimizations such as accumulation of multiple carry flags in a single register, and much timing-driven tuning of the code in order to yield further cycle-count reductions.

For large-integer arithmetic in the sense that the modulus $q$ is itself much larger than the natural integer wordsize of the underlying hardware, the cost of the algorithm is dominated by the two multiply operations, the MULL of which is a true half-multiply in that its cost is asymptotically half that of a general double-width product. The UMULH operation, as noted in \S~\ref{sect:modinv}, cannot generally be made similarly faster than a double-width product because we cannot be sure that the inputs are such that the carries into the high half of the ``multiplication rhombus" can be safely approximated by truncating the rhombus. However, as we shall show in the following section, the discarded low half of the full double-width product $tmp \cdot q$ here again has a simple closed-form expression which can be computed cheaply without any explicit multiplication, allowing the carries into the upper half to be accurately predicted and thus the UMULH result to again be computed at the same asymptotic cost as needed for the MULL, as was the case for the streamlined Newtonian mod-inversion. Thus we approximate the cost of each execution of the above loop for multiword moduli is asymptotically equal to that of two ``genuine half-multiplies", or equivalently a single general double-width product, ${\rm UMUL\_LOHI_b}(x, y)$.

We shall examine how to efficiently~-- that is, with asymptotically less effort than needed for the above loop, as $n$ becomes large~-- perform this scaling and final modular reduction step in \S~\ref{sect:algo_c}, after first considering a variant of the above remaindering algorithm, which will prove useful in terms of optimizing both the scaled-remainder computations, as well as the scaling step needed if the true remainder is desired.


\section{Fast Divisibility Check, Version 2}
\label{sect:algo_b}

This variant results from the seemingly trivial observation that since $q \cdot qinv \equiv 1\ ({\rm mod}\ R)$, if we modify the final step of the loop body in Algorithm~A by replacing the high-half multiply
$$
	cy = {\rm UMULH_b}(tmp, q);
$$
by a full double-width product which returns the lower and upper halves of $tmp \cdot q$ in a pair of b-bit integers:
$$
	hi:lo = {\rm UMUL\_LOHI_b}(tmp, q);
$$
then the value of the variable $lo$ on each loop execution simply recapitulates $x_i - cy (+ q)$, with the parenthetical (to denote that it is conditional) addition of $q$ occurring if a borrow resulted from the subtraction, that is, if $x_i < cy$. This allows the low half of the above double-width product to be cheaply computed without explicit multiplication, which in a multiword-operand setting permits a cheap form of error correction of the approximate ``probabilistic" carries into the the high-half of the product in a truncated-rhombus half-multiply implementation of the UMULH. This optimized variant is indicated in our pseudocode below via addition of $lo$ as an optional third argument to the UMULH operation.

Note that one case where this fails to offer a speed advantage is when $q$ fits into a hardware word and the compute architecture has no separate UMULH instruction. For example, an x86\_64-specialized version of Algorithm~A for 64-bit moduli has little choice but to use the above UMUL\_LOHI variant, since the x86\_64 MUL instruction automatically returns both halves of the 128-bit product in the RDX:RAX register pair, with RDX holding the high half of the product and RAX the low half. On a 64-bit architecture like the older (and groundbreaking for its time) DEC Alpha, on the other hand, one would prefer the separate-upper-half variant, since the Alpha ISA follows a strict ``two input register, one output register" RISC-style instruction paradigm and thus has separate hardware-multiply instructions (called MULQ and UMULH) to generate the low and high halves of a 128-bit unsigned product.


Of more immediate usefulness is the observation that the quantity $lo = x_i - cy (+ q)$ also satisfies a scaled-modular-partial-summation property, namely that on the $i$th loop iteration,
\begin{equation}
	lo = R^{-i} \sum_{j=0}^{i} a_j R^j ,\ {\rm for}\ i = 0, ... , n-1.
\label{Bsum}
\end{equation}
This allows us to cut one loop execution off of Algorithm~A, and replace the final loop execution of Algorithm~A with a simple multiply-free postprocessing step involving the high word of $x$. The resulting variant is captured in Algorithm~B.
\\
\begin{algorithm}
\SetAlgoLined
\SetAlgoRefName{B}
\vspace{0.1in}
\KwResult{True if $q$ divides $x$, false otherwise.}
{
	$uint_b\ tmp,\ lo,\ cy = 0$;\\
	\If {$n == 1$} {	{\em// Explicit mod needed in single-term case}\\
		\Return $x_0 \% q$\;
	}
	\For {$i\leftarrow 0$ \KwTo $n-2$}{
		$tmp  = x_i - cy$;\\
		{\em // Inline multiply-less computation of low half of tmp*q here,}\\
		{\em // for optional fast half-mul UMULH with carry correction:}\\
		$bw = (cy > x_i)$;\\
		\eIf {bw} {
			$lo = tmp + q$;\\
		}{
			$lo = tmp$;\\
		}
		$tmp = {\rm MULL_b}(tmp, qinv) + bw$;\\
		$cy = {\rm UMULH_b}(tmp, q [, lo])$;\\
	}
	{\em // Assume loop index is preserved: $i = n-1$. Final term needs no explicit multiplies:}\\
	$tmp  = x_i - cy$;\\
	\eIf {$cy > x_i$} {
		$lo = tmp + q$;\\
	}{
		$lo = tmp$;\\
	}
	\Return $(lo == 0)$\;
}
\caption{IS\_DIV\_B, right-to-left divisibility check, version 2. Data as for Algorithm A.}
\label{algo_b}
\end{algorithm}

%
%
If instead of merely checking divisibility of $x$ by $q$, we desire  the true remainder $r = x$ (mod $q$), one again simply uses a slightly modified functional interface (which we shall refer to as REMAINDER\_B but not give explicitly), whereby one returns the $b$-bit residue $lo$ directly, then takes the output value $lo$ and scales it as
$$
	r = lo \cdot R^{n-1} \ ({\rm mod}\ q).
$$
Note that unlike Algorithm~A, if $n = 1$, that is, the input has just a single term $x_0$, we must take care to do an explicit modular reduction of this term prior to returning, since otherwise $x_0$ passes through the loop unaltered into the output. This special case is easily handled as shown, however.

Table~3 lists the per-iteration values of the variable $lo$ (not computed explicitly in the above pseudocode) in the middle column for the same example computation as was used for Algorithm~A. It can be seen that the properly scaled~-- using~(\ref{Bsum})~-- values of this variable given in the rightmost column exactly match the scaled data in the analogous column of Table~2.
\begin{table}
\begin{center}
\label{table_b}
\caption{Per-loop-iteration data for Algorithm~B applied to $x = 2^{977}-1$ and $q = 16357897499336320049$, with $R=2^{64}$.}
\begin{tabular}{r|r|c}
	$i$	&	$lo$\qquad\qquad\qquad&$lo \cdot R^{i}$ (mod $q$)\qquad\qquad	\\
\hline
	 0	&18446744073709551615	&{\tt 1CFD12E467CEDBCE	}	\\
	 1	&10394635793536932813	&{\tt 4D611EA3809531E7	}	\\
	 2	& 5051339290092407161	&{\tt BD293142B725F2B8	}	\\
	 3	& 4156320137058648598	&{\tt 6DA6C745723D2042	}	\\
	 4	& 5752293599955516196	&{\tt 62A5DDA094A31338	}	\\
	 5	& 5424202733172914497	&{\tt 32D35CC447414DD0	}	\\
	 6	&10596859200044538054	&{\tt C4E358F892AFD7CB	}	\\
	 7	&18307282351603437371	&{\tt 841646B12435044D	}	\\
	 8	&11786040147344227072	&{\tt 953C002AC7E9CA9B	}	\\
	 9	& 5298951544689159434	&{\tt 87A198DD565BAE6E	}	\\
	10	& 5506369872691534183	&{\tt 665B982F2C57CF9F	}	\\
	11	& 9101239751444920986	&{\tt C2F464196442F72D	}	\\
	12	& 8007683592075709691	&{\tt 92AB735A1C3B4927	}	\\
	13	& 7266136640719894958	&{\tt C2A0BA67701C6754	}	\\
	14	&12039174030790701247	&{\tt 339D02265F21100D	}	\\
	15	& 4097145961007838330	&{\tt 77ABEA1607BF1817	}
\end{tabular}
\end{center}
\end{table}


\section{Efficient Powering to Obtain $R^n$ modulo $q$}
\label{sect:algo_c}

We next turn to computation of the scaling factors $R^n$ (mod $q$) and $R^{n-1}$ (mod $q$) required to compute the true remainder from the outputs of the Algorithm~A and B loops, respectively, that is to transform the outputs out of ``Montgomery space". To aid the discussion here we introduce a small piece of new notation, by defining $Mp(j,k)$ with $j,k$ nonnegative integers as the Montgomery product of the $j$th and $k$th powers of the radix $R$, yielding radix power $(j+k-1)$ (mod $q$):
$$
	Mp(j,k) := M(R^j\ ({\rm mod}\ q), R^k\ ({\rm mod}\ q), q; R) = R^{j+k-1}\ ({\rm mod}\ q) .
$$
We will also make use of an abbreviated form of this,
$$
	Mp(j,k) \ra (j+k-1) .
$$
Let us also define two specialized versions of the base-$R$ Montgomery modmul~-- Firstly a modular squaring $M(x,x)$:\\
\\
\begin{algorithm}[H]
\LinesNotNumbered
{\bf Algorithm:} {${\rm MONT\_SQR_b}$}\\
\KwData{Unsigned b-bit integers $x,q,qinv$, $q$ odd and $q \cdot qinv \equiv 1\ ({\rm mod}\ 2^b)$}
\KwResult{The radix $R = 2^b$ Montgomery square $x^2 \cdot R^{-1}$ (mod $q$)}
\vspace{0.1in}
{
	$uint_b\ hi:lo = {\rm USQR\_LOHI_b}(x)$;\\
	$lo = {\rm MULL_b}(qinv, lo)$;\\
	$lo = {\rm UMULH_b}(   q, lo)$;\\
	\eIf {$hi < lo$} {
		\Return $hi - lo + q$\;
	}{
		\Return $hi - lo$\;
	}
}
\label{MS}
\end{algorithm}
%

\vspace{0.1in}
\noindent And a ``downshift multiply", which effects a Montgomery multiply-by-unity, $M(x,1)$:\\

\begin{algorithm}[H]
\LinesNotNumbered
{\bf Algorithm:} {${\rm MMUL\_ONE_b}$}\\
\KwData{Unsigned b-bit integers $x,q,qinv$, $q$ odd and $q \cdot qinv \equiv 1\ ({\rm mod}\ 2^b)$}
\KwResult{The radix $R = 2^b$ Montgomery multiply by unity: $x \cdot R^{-1}$ (mod $q$)}
\vspace{0.1in}
{
	$uint_b\ lo = {\rm MULL_b}(qinv, x)$;\\
	$lo = {\rm UMULH_b}(   q, lo)$;\\
	\eIf {$lo \ne 0$} {
		\Return $q - lo$\;
	}{
		\Return $lo$\;
	}
}
\label{MUM}
\end{algorithm}
\vspace{0.1in}
%

For multiword moduli $q$ the USQR\_LOHI step in squaring-specialized version will be up to twice as fast as a general double-width product computed using UMUL\_LOHI, and the multiply-by-unity needs no double-width product at all.

The nature of the Montgomery modmul is such as to make the chief difficulty here ``getting off of square one", as it were. This is because one typically chooses the radix $R$ to be the smallest power of the machine wordsize $W$ such that $W^k > q$, making it relatively cheap to compute $R$ (mod $q$). For example, if $q < W = 2^{64}$ one could use the intrinsic 64-bit modulo function supported by one's favorite programming language, although this may or may not map to an efficient hardware instruction, and in practice is often very slow.

Once one has $R$ (mod $q$), if one's first instinct is to then perform a modular squaring, one now runs headlong into the fact that $Mp(1,1) \ra 1$ rather than the desired power, 2. One can of course start with $R$ (mod $q$) and compute $R^2$ (mod $q$) using a sequence of $b$ modular doublings. But especially if the required power $n$ is not terribly large it is worth trying to optimize the computation of this initial scaling factor $R^2$ (mod $q$). For example, if the integer-truncated ratio $\lfloor R/q \rfloor$ is sufficiently small as to be exactly storable in the 53 bits of an IEEE64 floating-point mantissa, one can compute the quotient $R/q$ using floating-point arithmetic (presumed to be faster than hardware integer division, as is true on most commercial CPU architectures), then cast back to integer and use integer arithmetic to compute $R$ (mod $q$) = $R - q \cdot \lfloor R/q \rfloor$. The author's own specialized code for the case $q < 2^{64}$ uses a small sequence of 64-bit floating point operations (including just one explicit floating-point inversion) to efficiently compute $2^{96} = R^{3/2}$ (mod $q$), then feeds that to ${\rm MONT\_SQR}_{64}$ to obtain $2^{128} = R^2$ (mod $q$). This approach can easily be generalized to the multiword-$q$ case.

In any event, once the starting scaling $R^2$ (mod $q$) is in hand we would like to use a minimal-cost sequence of the above two specialized modmul operations to compute $R^{n+1}$ (mod $q$) or $R^n$ (mod $q$), which are the radix powers needed to scale the Algorithm~A and B loop outputs, respectively, if we also use a Montgomery multiply for the scaling step. Our method is reminiscent of the classical algorithm used for left-to-right binary exponentiation~\cite{Knuth}, with a call to ${\rm MONT\_SQR_b}$ for each powering bit processed, this mod-squaring call being supplemented by a bit-parity-conditional call to ${\rm MMUL\_ONE_b}$. The latter ``downshift-multiplies" thus play the role of the scalar multiplies which augment the basic sequence of squarings in the left-to-right binary powering algorithm whenever a set bit is encountered in the exponent being left-to-right bit-scanned, with the added difference that in the present case, the added calls to ${\rm MMUL\_ONE_b}$ are triggered by zero bits rather than ones.

A further difference with classical powering ladders is that here we do not know {\em a priori} the needed sequence of operations, or at least we have not found a method to do so; however this is not problematic because we can use an inexpensive preprocessing computation to initialize a bitmap which is then used to control the subsequent powering sequence, by encoding the sequence of powers needed to obtain the desired power of $R$ (mod $q$) via calls to the above two specialized versions of the modmul. The preprocessing loop works backward from the final power, roughly halving (halving-plus-one, to be precise) the current power at each step and using its parity to encode the needed operation in the resulting bitmap.

In order to simplify the discussion, let us assume we are using Algorithm~B, thus that we need $R^n$ (mod $q$) as the radix-power input to the output-scaling Montgomery multiply used to produce the true remainder $x$ (mod $q$). Assuming $n$ to be at least 3, starting with power $p = 2$ we always make one immediate call to ${\rm MONT\_SQR_b}$, yielding power 3. For $n > 3$ we now have the following 2 choices:
\vspace{-0.1in}
\begin{align}
\nonumber	{\bf [a]}\qquad Mp(2,3) &\ra 4;	\qquad\qquad	\\
\nonumber	{\bf [b]}\qquad Mp(3,3) &\ra 5,	\qquad\qquad
\end{align}
which differs from subsequent iterations, for which our two options (based on the input power $p$ at that iteration) are as follows:
\vspace{-0.2in}
\begin{align}
\nonumber	{\bf [c]}\qquad &Mp(p,p) \qquad\ \ \ \ \ra 2p-1;	\\
\nonumber	{\bf [d]}\qquad &Mp(p,Mp(p,0)) \ra 2p-2.
\end{align}
The preprocessing step does this by working backwards and encoding the needed steps as a sequence of bits:
\begin{align}
\nonumber	&{\rm Bit}\ j = 0 :\qquad 0|1\ {\rm means}\ {\bf [a]}\ {\rm or}\ {\bf [b]},\ {\rm respectively};	\\
\nonumber	&{\rm Bit}\ j > 0 :\qquad 0|1\ {\rm means}\ {\bf [c]}\ {\rm or}\ {\bf [d]}\ .
\end{align}

Since the {\bf [a]}/{\bf [b]}-choice can be made based on the parity of the power $p$ resulting from the halving loop, this frees up a bit; accordingly in our implementation we initialize the bitmap bit-index $j$ to 0 and use it strictly for the {\bf [c]}/{\bf [d]}-choice selection control. The Algorithm~C pseudocode illustrates this powering scheme. The `bitmap' type here is simply shorthand for a suitably long integral type to hold the string of resulting bits. Unless the power $n$ is truly gargantuan, meaning $x$ larger than several gigawords, a 32-bit integer is perfectly adequate to the task.

\textheight=9.5in

\begin{algorithm}
\SetAlgoLined
\SetAlgoRefName{C}
\KwData{Array size $n > 2$, $uint_b\ pow \equiv R^2$ (mod $q$), $q,qinv$, $q$ odd and $q \cdot qinv \equiv 1\ ({\rm mod}\ 2^b)$}
{
	\eIf {${\rm n} == 3$} {
		\qquad\ \ \ $pow \leftarrow {\rm MONT\_SQR_b}(pow,\ q,\ qinv)$;	\qquad{\em// pow  = $R^3\ ({\rm mod}\ q)$}\\
	}{
		$uint_b\ ptmp \leftarrow {\rm MONT\_SQR_b}(pow,\ q,\ qinv)$;	\qquad{\em// ptmp = $R^3\ ({\rm mod}\ q)$}\\
		$int\ j \leftarrow 0,\ p \leftarrow n, bm \leftarrow 0$;	\qquad	{\em// Init loop parameters and bitmap}\\
		\While{$p > 5$} {
			{\em// Each M-mul accumulates another inverse power 1/R; add 1 to}\\
			{\em// current power after each halving step here to account for this:}\\
			${\rm BIT\_SETC}(bm,\ j,\ {\rm IS\_EVEN}(p))$;\\
			$p \leftarrow (p/2) + 1$;\qquad $j \leftarrow j + 1$;\\
		}
		{\em// Next step depends on value of p resulting from above loop:}\\
		\eIf {${\rm p} == 4$} {
			$pow \leftarrow {\rm MONT\_MUL_b}(pow,\ ptmp,\ q,\ qinv)$;\\
		} ( {\bf if}\ ${\rm p} == 5$\ {\bf then}) {
			$pow \leftarrow {\rm MONT\_SQR_b}(ptmp,\ q,\ qinv)$;\\
		}
		{\em// Bitmap-controlled powering loop is only entered if $n > 5$:}\\
		\For {$i\leftarrow j-1$ \KwTo $0$}{
			\eIf {${\rm BIT\_TEST}(bm,\ i)$} {
				$ptmp \leftarrow {\rm MMUL\_ONE_b}(pow,\ q,\ qinv)$;	{\em// Reduce R-power by 1}\\
				$pow  \leftarrow {\rm MONT\_MUL_b}(ptmp,\ pow,\ q,\ qinv)$;	{\em// ...and multiply.}
			}{
				$pow  \leftarrow {\rm MONT\_SQR_b}(pow,\ q,\ qinv)$;
			}
		}
	}
	\Return $pow$\;
}
\caption{RADIX\_POWER, computes $R^n$ (mod $q$)}
\label{POW}
\end{algorithm}
%
%
%
The function BIT\_SETC$(bitmap\ bm,\ int\ j,\ boolean\ cond)$ here sets the $j$th bit of the bitmap argument if the condition $(cond)$ is true; in our case, if the current value of the power $p$ is even. The reason we only set any bits for input powers $> 5$ is because smaller powers are handled by the special-casing described above to obtain $p = 4$ or 5 from the pair of starting powers $p = 2$ and 3. (Note that we could start things off with just the single power $p = 3$ and then use the option {\bf [c]}/{\bf [d]} switch to obtain either $p = 4$ via $Mp(3,Mp(3,0))$ or $p = 5$ by way of $Mp(3,3)$ but as we have already used $p = 2$ to get $p = 3$ it is quicker to use the power $p = 2$ directly rather than recomputing it as $Mp(3,0)$).

\textheight=9in

The final step is to use the resulting power to multiply the scaled remainder resulting from the mod-loop. Depending on whether one computes a remainder uaing Algorithm~A or B, one feeds either $n+1$ or $n$ to the radix-powering step:
\begin{align}
\nonumber	&// Algorithm\ A\ flow:		\qquad\qquad\qquad\qquad\ \ \ \ \ \ // Algorithm\ B\ flow:\\
\nonumber	&tmp = {\rm REMAINDER\_A}(x,q,qinv,n);	\qquad\ \ \ \,	tmp = {\rm REMAINDER\_B}(x,q,qinv,n);\\
\nonumber	&pow = {\rm RADIX\_POWER}(q,qinv,n+1);	\qquad\,	pow = {\rm RADIX\_POWER}(q,qinv,n);
\end{align}
\noindent In both cases one then computes the true remainder $x$ (mod $q$) via $r = {\rm MONT\_MUL_b}(tmp,pow,q,qinv).$


It is useful to have both options because it affords another modest optimization opportunity, namely that of precomputing the powering bitmaps for both $n$ and $n+1$ (that is, just the BIT\_SETC~-~containing loop above) prior to performing the main bitmap-controlled powering loop. One can then take the result with the fewest set bits to generate the scaling needed to apply to the output of Algorithm~A or B, respectively. Since $n$ is given, this can all be done at the outset and then used to choose which scaled-remainder algorithm to use. The general rule is: If the even element of the $(n,n+1)$ pair has fewer set bits than the odd we should use Algorithm~A for the remainder computation, otherwise we should use Algorithm~B. The borderline case of the even element having one less set bit is a wash, because the cost of the extra loop execution and concomitant MULL~/ UMULH pair of Algorithm~A offsets the cost savings of one fewer call to the MMUL\_ONE function.

Let us again illustrate with a small numerical example: For our example of $x = 2^{977}-1$ and $q = 16357897499336320049$, if using the Algorithm~B output the radix-powering step begins with $n = 16$, and the bitmap-creation loop produces successive values $p = 16$ and 9, at which point it exits with a binary bitmap $bm = 01$ and $p = 5$. Beginning with the precomputed square $R^2 = 5575771501247148520$ (mod $q$), the algorithm then does the sequence of four steps summarized in Table~4 with resulting mod-powers of $R$, the first two of which are done by the special-casing in the above code and the last two by the final for-loop in Algorithm~C, which runs through the bitmap in reverse bit order.
\begin{table}[ht]
\begin{center}
\label{r16}
\caption{RADIX\_POWER steps to compute $R^{16}\ ({\rm mod}\ q)$}
\begin{tabular}{r|r|l|l|r}
	pow\_in	&bit	&function calls			&operation			&pow\_out	\\
\hline
	 2		&  -	&MONT\_SQR				&Mp( 2, 2)			& 3			\\
	 3		&  -	&MONT\_SQR				&Mp( 3, 3)			& 5			\\
	 5		&  0	&MONT\_SQR				&Mp( 5, 5)			& 9			\\
	 9		&  1	&MONT\_SQR, MMUL\_ONE	&Mp( 9, Mp(9,0) )	&16
\end{tabular}
\end{center}
\end{table}

\vspace{-0.2in}
This yields $R^{16} = 1547775041475743422$ (mod $q$), and a final Montgomery-multiply of this with the REMAINDER\_B output $tmp = 4097145961007838330$ yields the true mod, $r = 8623243291871090711 \equiv 2^{977}-1$ (mod $q$).  If we instead use the Algorithm~A route we execute the radix-powering step beginning with `n' (really $n+1$) = 17, the bitmap-creation loop produces successive values $p = 17$ and 9 and exits with bitmap $bm = 00$ and $p = 5$. For this exponent the resulting powering sequence requires no Montgomery-mul-by-unity, as shown in Table~5.
\begin{table}[ht]
\begin{center}
\label{r17}
\caption{RADIX\_POWER steps to compute $R^{17}\ ({\rm mod}\ q)$}
\begin{tabular}{r|r|l|l|r}
	pow\_in	&bit	&function calls			&operation			&pow\_out	\\
\hline
	 2		&  -	&MONT\_SQR				&Mp( 2, 2)			& 3			\\
	 3		&  -	&MONT\_SQR				&Mp( 3, 3)			& 5			\\
	 5		&  0	&MONT\_SQR				&Mp( 5, 5)			& 9			\\
	 8		&  0	&MONT\_SQR				&Mp( 9, 9)			&17
\end{tabular}
\end{center}
\end{table}

The result is $R^{17} = 8502984233828494641$ (mod $q$), and a final Montgomery-multiply of this with the REMAINDER\_A output $tmp = 11326687669760783497$ again yields the desired true mod.

In order to collect some statistics about the efficacy of counting the bitmap set-bits for such pairs and then choosing Algorithm~A or B depending on the result, we have calculated the number of set bits in the bitmap for all $(n,n+1)$ pairs from $n = 6$ (the smallest value for which Algorithm~C creates a bitmap) up to $n = 2^{20}$. The average difference in the number of bits is just over 1 (we obtain 1.0000123948), but the maximal difference is asymptotically the same as the number of bits in the bitmap. This can be shown by examining the worst-case scenario, which is when $n = 2^a + 1$, for which $n$ yields a bitmap of $(a-1)$ binary zeros and $n+1$ yields one of $(a-1)$ binary ones. Again, this is only germane if $n$ is not huge, due to $O(\lg n)$ modmul count of Algorithm~C, which is asymptotically vanishing relative to the cost of the remaindering loops of Algorithms A and B as $n\ra\oo$.


\section{Use of the Remainder to Obtain the Quotient}
\label{sect:algo_d}

The quotient-producing algorithm again starts with the basic property of the Montgomery inverse,
$$
	q \cdot qinv \equiv 1\ ({\rm mod}\ R).
$$
The general theme here is expressed by the question ``can we multiply the input $x$ by the mod-inverse $qinv$ to somehow obtain the quotient?" Starting with the original dividend $x$ this leads nowhere, but once we have computed the remainder $r = (x\ {\rm mod}\ q)$, we note that the difference $(x - r)$ is divisible by $q$, that is, $(x - r) = k \cdot q$, with the quotient $k$ = $\lfloor x/q \rfloor$ a nonnegative integer. Thus
$$
	(x - r) \cdot qinv \equiv k\ ({\rm mod}\ R) .
$$
In other words ${\rm MULL_b}(x - r, qinv)$ yields the bottom $b$ bits of the quotient $k$, exactly. If $k < R$ (that is, $x < (q + 1)R$) this is all we need. For arbitrarily large quotients, that is, $k$ possibly much larger than the modulus $q$, one approach is to ``extend the mod-inverse" by doing more Newton iterations to produce a ``full-sized" mod-inverse $qinv_n$, such that
$$
	q \cdot qinv_n \equiv 1\ ({\rm mod}\ R^n),
$$
for which a length-$R^n$ ${\rm MULL}(x - r, qinv)$ yields the full quotient, exactly. But this is very wasteful in that the work scales supralinearly in $n$. At this point we take a clue from the remainder computation and realize that we can combine the above idea with the right-to-left Montgomery-mul-based carry scheme to compute the quotient in linear-work fashion, one radix-$R$-sized piece at a time. The resulting method is captured in Algorithm~D. The loop here simply recapitulates the Hensel-remainder loop of Algorithm~A, but by initializing the value of the loop-carry $cy$ to equal the remainder we now apply the algorithm to $(x - r)$, which is guaranteed to have remainder (both true and Hensel) 0. The carry yielded by the UMULH output of each loop execution effectively "unwinds" the Hensel-remainder summation, one term at a time, and the preceding MULL produces the corresponding quotient term.

With respect to the possibility of integer overflow in the carry/borrow sum subtracted from the temporary in the first step of the loop body, since values of $cy$ subsequent  to the initializer are the result of a UMULH, they cannot exceed the largest possible UMULH output, UMULH$(R-1, R-1) = \lfloor (R^2 - 2R + 1)/R \rfloor = R-2$, thus permitting the subtraction of both $cy$ and the binary (strictly 0 or 1) borrow resulting from the preceding loop execution in the first step of the loop body. (In other words the sum $cy + bw$ cannot overflow). The modmul portion of the loop body is more or less identical to that of Algorithms A and B, with the crucial difference that the output of each loop execution consists of two $b$-bit data: the current word of the quotient, which is in the MULL output, and the leftward carry into the next-higher word, which is the UMULH output. No radix-power scaling of the output is needed.
\vspace{0.1in}

\begin{algorithm}[H]
\SetAlgoLined
\SetAlgoRefName{D}
\KwData{As for Algorithm~A, plus the remainder $r$ and an integer array $y[n]$ to take the result.}
\KwResult{The quotient $\lfloor x/q \rfloor$ is returned in $y$, which may coincide with $x$. On output, $cy=0$.}
\vspace{0.1in}
{
	$uint_b\ tmp,\ bw = 0,\ cy = r$;\\
	\For {$i \leftarrow 0$ \KwTo $n-1$}{
		$tmp = x_i - bw - cy$;\\
		$bw = (tmp > x_i)$; {\em// Unsigned compare of result to check for a borrow}\\
		$tmp = {\rm MULL_b}(tmp, qinv)$;\\
		$cy = {\rm UMULH_b}(tmp, q)$;\\
		$y_i = tmp$; {\em// Write current quotient word, which is in the MULL output}\\
	}
}
\caption{The quotient computation}
\label{algo_d}
\end{algorithm}

%

There is also a key difference between Algorithms A/B and Algorithm~D with regard to the subtraction which begins each loop execution. Here, since the quotient computation is literal rather than $({\rm mod}\ q)$, if the subtraction results in a borrow, that is accounted for by carrying the resulting $-1$ into the next-higher word. Similarly to the remainder computation, the above quotient algorithm allows for a fast computation of the (discarded) low half of the final $tmp \cdot q$ product, which could be used in a streamlined error-corrected version of a multiword UMULH: Here the low half of a full double-width ${\rm UMUL\_LOHI_b}(tmp, q)$ must simply be equal to the result of the opening $tmp = x_i - bw - cy$ operation\footnote{Aug 2016: Previous versions of this manuscript contained an unfortunate extraneous $+ bw$ following the MULL in the loop body pseudocode listing, due to a lazy copy-paste-modify from Algo A on the part of the author. This differed both from the author's C code implementation and the 2-folded assembly code listing in Table~6. Thanks to Jens Nurmann for catching this.}.

For our example $x = 2^{977}-1$ and $q = 16357897499336320049$, Table~6 lists successive $y_i$ terms for the above loop, which can easily be verified as the coefficients of the quotient $\lfloor x/q \rfloor$ in base-$2^{64}$ arithmetic.
\begin{table}
\begin{center}
\label{table_d}
\caption{Per-word data for Algorithm D applied to $x = 2^{977}-1$ and $q = 16357897499336320049$, with $R=2^{64}$. Right column: An x86\_64 assembly-code implementation of the 2-way-folded quotient loop. Execution order matches natural reading order.}
\begin{tabular}{r|r| l}
i	&Quotient word $y_i$\qquad\qquad	&	\\
\hline
	 0	& 6364180061714936936	&	{\tt // RCX=n/2, RSI=q, RBX=qinv, RDI,RDX=cy0,1, R8,R9=bw0,1=0}\\
	 1	& 4771973621301622518	&	{\tt	LEA	\ (R10,RCX,8),R11	\qquad	// R11 = \&x+n/2 (\&x in R10)}\\
	 2	&  694724920058399436	&	{\tt	LEA	\ (R13,RCX,8),R14	\qquad	// R14 = \&y+n/2 (\&y in R13)}\\
	 3	& 7462732776264284083	&	{\tt loop:						\qquad	// Loop from n/2 to 1}\\
	 4	&15651191667900344027	&	{\tt	SUB \ R8 ,RDI			\qquad	SUB \ R9 ,RDX	// (bw+cy) via cy-(-bw)}\\
	 5	&  684779273839653350	&	{\tt	MOV \ (R10),RAX			\ \ 	MOV \ (R11),R12	// x[i], x[i+n/2]}\\
	 6	& 8910056920539811989	&	{\tt	SUB \ RDI,RAX			\qquad	SBB \ R8 ,R8 	// tmp0=x0-(bw0+cy0), R8=-CF}\\
	 7	& 6625598233439971816	&	{\tt	SUB \ RDI,RDI			\qquad	SBB \ R9 ,R9 	// tmp1=x1-(bw1+cy1), R9=-CF}\\
	 8	&13578887251066731535	&	{\tt	ADD \ 8,R10				\qquad\ \ 	ADD \ 8,R11		// Increment x-array pointers}\\
	 9	& 7249027741998019233	&	{\tt	IMUL\ RBX,RAX			\qquad	IMUL\ RBX,R12	// tmp0,1 *= qinv}\\
	10	&11772736962114281085	&	{\tt	MOV \ RAX,(R13)			\ \ 	MOV \ R12,(R14)	// y[i],y[i+n/2] <-- tmp0,1}\\
	11	&15530135107470554958	&	{\tt	ADD \ 8,R13				\qquad\ \ 	ADD \ 8,R14		// Increment y-array pointers}\\
	12	& 6468054066637286049	&	{\tt	MUL \ RSI		\qquad	\qquad	MOV \ RDX,RDI	// cy0 = MULH64(q,tmp0)}\\
	13	& 8083046564352798341	&	{\tt	MOV \ R12,RAX			\qquad	MUL \ RSI\ \ \ \ \ // cy1 = MULH64(q,tmp1)}\\
	14	&              147809	&	{\tt DEC RCX	}\\
	15	&                   0	&	{\tt JNZ loop					\ \ 	// loop if RCX != 0}\\
\end{tabular}
\end{center}
\end{table}
%

At this point we feel compelled to make a few more comments regarding the building-up of the quotient in the above method. We illustrate things using a slightly smaller example, $x$ fitting into three 64-bit words and the same 128-bit modulus $q$ we used in Section 1 to illustrate the streamlined mod-inverse computation:
\begin{align}
\nonumber&	x = 153238840814299457340643142885404331762436489574620087,\ q = 225797717267637708506527464987314161,\\
\nonumber&	qinv = 98317950452290864966529955359911823633 = 5329826773734796952 \cdot 2^{64} + 18061898331188349201\ .
\end{align}
Using $R = 2^{128}$ the Algorithm~A loop output is 204005585172405248723488909070772720, which is negated and multiplied by $R^2$ (mod $q$). The Algorithm~B loop output is 85991941466095605134529253504604205, which is multiplied by $R$ (mod $q$). Both methods obviously give the same remainder,
$$
	r = 130392762589805994888402779408669015\ .
$$
We next form $(x - r)$ in preparation for the quotient computation:
\begin{align}
\nonumber	&x - r = 153238840814299457210250380295598336874033710165951072\\
\nonumber	&= 450328479259411 \cdot 2^{128} + 18248253616373032484 \cdot 2^{64} + 17701223841397244512\\
\nonumber	&= 450328479259411 \cdot 2^{128} + 336620864253378130591640020431239938656\ .
\end{align}
MULL of the bottom 128 bits of $(x - r)$ with $qinv$ gives the exact quotient $y = \lfloor x/q \rfloor = (x - r)/q$:
$$
	{\rm MULL_{128}}(336620864253378130591640020431239938656, qinv) = 678655403024582752\ .
$$
At this point we can skip the remaining Algorithm~D loop execution if we like, since it only serves to nullify the remaining bits of $(x - r)$. This, and the fact that the loop runs through the words of $(x - r)$ in strictly ascending order, highlights a seemingly curious property of the quotient computation: we use only the $j$th and lower words of $(x - r)$ to compute the $j$th word of the quotient $y$. Thus, if we compute in advance the expected number of nonzero words in $y$, we can compute $y$ using only that number of low words of $(x - r)$. In the above example, since we can see easily that the quotient will fit in a 64-bit word, we can exit the loop early as noted, but we can in fact use just the low 64 bits of $qinv$ and compute the quotient using just a single MULL of that with the the low 64 bits of $(x - r)$:
$$
	{\rm MULL_{64}}(17701223841397244512, 18061898331188349201) = 678655403024582752\ .
$$
The idea that one can compute the exact quotient without explicit reference to the high words of $(x - r)$ may strike readers whose intuition in such matters has been shaped by classical long division (that is, most of us) as bizarre, but note that the datum $(x - r)$ is the result of parsing the entire original input $x$ being divided in order to compute the remainder; the Montgomery modmul-based quotient algorithm is simply exploiting the fact that the remainder-subtraction has in effect ``shifted" all the information needed to compute the quotient $y$ into the bottom $\lceil \lg y \rceil$ bits of the quantity $(x - r)$. (We admit that seeing this effect in action can be somewhat startling at first.)

We have not found it possible to modify the above sequence to permit for obtaining of the remainder and the quotient without the above second pass through the (remainder-subtracted) data, that is, ``in one fell swoop" as is done in classical long divisions implemented in left-to-right fashion. However, the quotient algorithm permits the same kind of folded-loop interleaved-instruction-stream optimization as we described for the remaindering algorithms (using Algorithm~A for the detailed demonstration), when $x$ is large relative to the divisor $q$. There is a small twist when applying this approach to the quotient, however, which we illustrate using the smallest nontrivial folding factor, $F=2$. The quotient Algorithm~D applied to the resulting 2 half-length vectors now needs partial remainders $r_0$ and $r_1$ which will be subtracted from the lower and upper dividend halves prior to entering the algorithm's loop. The upper-half partial remainder is just that, $r_1 = \lfloor x/R^{n/2} \rfloor$ (mod $q$). But for the lower-half portion of the computation, we consider again a concrete case such as captured in Table~6, and note that the lower-half quotient data must be the same as those produced by the serial version of the algorithm, e.g.~quotient words $y_0-y_7$ in the table. Thus the correct lower-half partial remainder is {\em the full-dividend remainder}. The generalization of this to larger folding factors involves a ``downward cascade" of partial remainders, in which each partial remainder is that for the entire portion of dividend ($x$) array above it. For instance, if we are using a loop-folded implementation of Algorithm~A for the remaindering, we can compute the needed terms via a slight rearrangement of the partial-remainders in~(\ref{algo_a_part}). Rather than using the nesting~(\ref{algo_a_nest}), which is inappropriate if the quotient is being computed because the intermediate partial remainders are negated (mod $q$), we now explicitly negate each term to produce the required partial remainders in descending, chained fashion:
\begin{equation}
r_{F-1} = -cy_{F-1}\cdot P,\ r_{F-2} = (r_{F-1}-cy_{F-2})\cdot P,\ ...,\ r_0 = (r_1-cy_0)\cdot P,
\label{algo_d_part}
\end{equation}
where all terms are computed (mod $q$). This simply reflects the "unwind" of the Hensel-remainder summation in a periodic-checkpoint form matched to the parallel remaindering step used to compute said remainder. The observant reader will be asking how to account for the ``missing borrows" here, e.g.~for $F=2$ if the processing of the $i=n/2-1$ term in Algorithm~D leaves us with $bw=1$, which in the serial version of the algorithm is subtracted from the low word of the upper half of the $x$-vector, $x_{n/2}$. In loop-folded mode this portion of the vector starts with its borrow set to 0, but note that this automatically compensated for in the above construction of the partial remainders, since the resulting carry subtracted from the $x_{n/2}$ term will be one larger as a consequence of the same truncation of the borrow chain in the loop-split remaindering computation.

For the full-DIV (2-pass, one for remainder and the second for quotient) algorithm, again using GCC on the Core 2 architecture, we see an appreciable speed difference between our C code and assembler implementations of the 2-way and 4-way-split algorithm, likely because the need to compute distinct array read and write addresses~-- we used an out-of-place implementation for generality, rather than making use of the expedient of overwriting the dividend by the quotient~-- leads to severe register pressure even in the 2-way-split case, as our x86\_64 assembly code for the loop body in the rightmost column of Table~6 illustrates. This is a direct pseudocode rendering of our inline assembler, and 13 of the available 14 general-purpose registers (RSP and RBP are reserved for use by the operating system and compiler) are used. For the 2-way-split case, our C code needs 21 cycles per input word for the full DIV; our assembler version needs just 17 cycles. In the 4-way-split case the C code needed 15.5 cycles per word, compared to a mere 12.5 for the assembly. Thus, for a register-lean custom assembler implementation of the 2 and 4-way-split quotient loops, we obtain virtually identical timings for the quotient computation as for the remaindering.

One other note about the assembly coding, as illustrated by the example snippet, relates to the issue of detailed scheduling of the high-latency IMUL and MUL instructions. The MUL is especially interesting because it has both high latency and the x86-designer-imposed constraints of the implicit input in RAX and the two output words being written to the RAX and RDX registers\footnote{This user constraint is alleviated in the next-generation x86 architecture, ``Haswell" via a 3-operand MULX instruction.}. Thus if one is doing more than one MUL in a given block of assembly, one must copy the outputs which are needed (in our case RDX) to a different register or to memory, then load one datum of the following MUL input pair into RAX before doing the next MUL. One would think that this would make for an instruction scheduling nightmare, but that is not the case. Notice in our assembly that we appear to blithely disregard the timing impact of the foregoing register-copy issues and simply schedule the MULs and succeeding output-copying MOV instructions in simple sequential order (lines $i=12$ and $i=13$ of the table). This is because the x86 has many more physical registers than the mere 16 named ones visible to the programmer. This large ``hidden" register pool supports a very powerful runtime-optimization capability, via the register-rename and out-of-order execution engines built into the processor. These runtime optimizations will typically assign separate physical registers to store such (apparently) troublesome intermediates, although it is impossible to be certain without a cycle-accurate simulator of these chip internals. Nonetheless, our timings provide indirect evidence of this process at work, since the resulting timings would not be possible if the dependencies were real.

Thus, the dependence of the instruction sequence as it relates to getting MUL inputs into the RAX register and outputs out of RAX and RDX is real as far as register availability for the coding is concerned, but is false with respect to the resulting data movement actually occurring or having runtime impact, and as an assembly programmer (or compiler-optimizer writer) one must be aware which scheduling constraints are real, and which are not. This neatly captures a key theme related to code optimization for such heavily virtualized execution engines: We have found that the biggest performance gains are achieved by giving the processor a reasonably full menu of independently executable instructions at all times and then ``just getting out of its way", so to speak, rather than trying to micromanage instruction streams which will inevitably be rescheduled at runtime in a fashion which is invisible to, and hence beyond the control of, the programmer.

By way of comparison with a state-of-the-art implementation of left-to-right DIV, our current-best 12.5 cycles per dividend word compares to 25 cycles for the current GMP release on the same architecture~\cite{GMP_div} (cf.~\S~V, the case study of the DIV\_NBY1 implementation)\footnote{On Haswell, the unmodified 4-way-split loop inline-assembly code which yielded the best results on Core2 runs in 3.0 and 6.8 cycles/word for remainder-only and full-div, respectively. We expect similar cycle counts on the earlier Sandy/Ivy Bridge CPUs, as the main boost is due to the significantly improved throughput of the integer MUL instructions already implemented in these of Intel's post-Core2 chip families~\cite{agner}. The x86 implementation of the algorithm has no interactions between the MUL instructions of the loop body and the add/sub carry flags, thus modifying the code to use the new-in-Haswell MULX instruction is only of modest benefit, in that it obviates some register-data moves related to the old x86 MUL ax:dx register-pair constraints; the resulting cycle counts are 2.8 and 6.6 cycles/word for remainder-only and full-div, respectively. As on Core2, these are roughly double the throughout of the analogous GMP functions running on the same architecture~\cite{GMPtiming}}. It is possible that classical left-to-right DIV algorithms (which the GMP $n/1$ division implementation represents, albeit in a very cleverly structured and highly optimized form) may benefit from the same kind of loop-folding optimization we describe here. However, were such a parallelized approach both possible and advantageous for the left-to-right DIV, one would expect it to have been employed some time ago, since latency issues related to integer MUL are not a new phenomenon. In fact, note that the 2-pass separate remainder/quotient structure of our approach here is key to this loop-folding parallelization, since the initial remaindering pass provides the partial-vector remainders needed to perform the quotient step in analogous parallel fashion. Time will tell whether classical left-to-right DIV can also benefit from a similar kind of optimization.

Lastly, what to do if $q$ is even? That is quite simple: One starts as described in the analogous Algorithm~A commentary, by counting the trailing binary zeros of $q$ and saving that value in an integer variable $tz_q$. Then one does as follows:
\vspace{0.1in}

\begin{algorithm}[H]
\SetAlgoLined
\SetAlgoRefName{D'}
\begin{enumerate}
\item	Save the bottom $tz_q$ bits of $x$: $bsave = x\ \& (2^{tz_q} - 1)$;
\\
\item	Right-shift $q$ and $x$ to obtain $q' = (q \gg tz_q)$ and $x' = (x \gg tz_q)$;
\\
\item	The combination of Algorithm~A/B and C applied to $x'$ and $q'$ yields a partial remainder $r'$,\\
from which the true remainder is obtained via $r = (r' \ll tz_q) + bsave$;
\\
\item	Algorithm~D applied to $x'$, $q'$ and $r'$ yields the quotient, without any other adjustment required.
\end{enumerate}
\caption{Modifications to Division procedure needed for even divisor $q$}
\end{algorithm}
\vspace{0.1in}


\section{Efficient Inverse Power of 2 Modulo q}
\label{sect:algo_e}

We conclude by considering the special case $x = 2^p + a$. Computing $x$ (mod $q$) here calls for a binary powering ladder, e.g.~a left-to-right binary exponentiation starting with seed $s = 2$ corresponding to the leftmost set bit of the exponent $p$, and parsing the lower-order bits of $p$ in rightward fashion. For each bit along the way one squares the current residue $s$ (mod $q$) and if the current bit $= 1$ one further performs a mod-doubling of the result. At the end one can check divisibility via $s \equiv -a$ (mod $q$), or compute the full remainder $r \equiv s + a$ (mod $q$). Thus, only $\lfloor \lg(p) \rfloor$ modmuls are required.

If using a radix-$R$ Montgomery modmul for this purpose one typically scales the initial seed as $s' = s \cdot R$ (mod $q$). Each mod-squaring then is replaced by a scaled variant
$$
	M(s', s') = M(s R, s R) \equiv s^2 R^2 R^{-1} \equiv s^2 R\ ({\rm mod}\ q) ,
$$
thus the scaling factor $R$ is carried through the sequence of squarings in constant fashion, and we note the bit-dependent mod-add does not alter this property, due to the distributivity of modular addition. Thus one can perform the powering as usual and at the end transform the output via modmul with $R^{-1}$ (mod $q$), which can be effected with a single call to ${\rm MMUL\_ONE_b}$.

However, because the dividend involves a power of 2, as does the radix $R$, we can make the inverse power of $R$ introduced by each modmul work to our advantage, and thus entirely avoid the need for the explicit $R$-scalings. Using that $R = 2^b$ we define a modified power
$$
	p' = p + b,
$$
which is used to control the mod-doublings in the powering sequence as described above, with one twist: Since we will be computing a sequence of negative powers culminating in $2^{-p}$ (mod $q$), we do a mod-doubling whenever the current bit of $p'$ is 0, or as in the pseudocode below, whenever the bit of of the logical complement $pshift := {\rm NOT}(p')$ is set. As is the case in normal (positive-exponent) binary powering, we can save some work by setting our initial seed not to 2 but rather to $2^x$, where $x$ is the largest chunk of bits of $p'$ starting with the leftmost set bit which is less than $b$, the power of 2 in our arithmetic radix. For simplicity we here assume that $p$ and $q$ are both $< R$, although it is easy to accommodate arbitrarily large exponents. Within a $uint_b$ datum, bits are indexed as [b-1:0], from left to right:
\vspace{0.1in}

\begin{algorithm}[H]
\SetAlgoLined
\SetAlgoRefName{E}
\KwData{Unsigned $b$-bit exponent $p$, $q$ odd, $q \cdot qinv \equiv 1\ ({\rm mod}\ 2^b)$.}
\KwResult{$2^{-p}$ (mod $q$).}
\vspace{0.1in}
{
	$uint_b\ pshift = p + b$;\qquad{\em// Scale p by power of 2 in the arithmetic radix}\\
	$int\ leadz  = ${\em [number of leading zero bits in pshift];}\\
	$int\ i_1  = b - 1 - leadz$;\qquad{\em// Index of leftmost set bit in pshift.}\\
	$int\ i_0 = ${\em [Starting at $i_1$, smallest bit index such that the bitfield $[i_1:i_0] < b$]};\\
	$int\ ichunk = (pshift \gg i_0)$;\qquad\qquad{\em// This\ is\ the\ bitfield\ $[i_1:i_0]$}\\
	$uint_b\ s = 1 \ll (b - ichunk - 1)$;\qquad{\em//\ The\ resulting\ initial\ seed}\\
	$pshift = {\rm NOT}(pshift)$;\qquad{\em//\ Logical\ complement\ of\ the\ scaled\ exponent}\\
\vspace{0.1in}
	{\em//\ Loop\ downward\ though\ the\ bits\ of\ the\ scaled,\ complemented\ exponent}\\
	\For {$int\ i \leftarrow i_0-1$ \KwTo $0$}{
		$s = {\rm MONT\_SQR_b}(s,\ q,\ qinv)$;\\
		\If {${\rm BIT\_TEST}(pshift,\ i)$} {
			$s = s + s$;\\
			${\rm if}(s \ge q)\ s = s-q$;\\
		}
	}
	{\em// Always need a final mod-doubling prior to returning.}\\
	$s = s + s$;\\
	${\rm if}(s \ge q)\ s = s-q$;\\
	\Return $s$;
}
\caption{Computation of $2^{-p}$ (mod $q$) without any explicit radix-scalings.}
\label{algo_e}
\end{algorithm}
\vspace{0.1in}

%
%
%
We illustrate once again using our earlier example, the problem of finding the remainder of $x = 2^{977}-1$, modulo $q = 16357897499336320049$. The 64-bit mod~-inverse $qinv = 9366409592816252113$. Working through the steps of Algorithm~E, we have $pshift = 977 + 64 = 1041$, which in binary form
$= 10000010001_2$. Using $R = 2^{64}$, we have $leadz = 53$ and $i_1 = 64-1-53 = 10$. The leftmost 6-bit-wide chunk of $p$ is guaranteed to be in $[32, 63]$, hence $i_0 = i_1-6+1 = 5$, and the 6-bit-wide bitfield in question is $ichunk = (p \gg 5) = 100000_2 = 32$. The resulting initial seed $s = (1 \ll (64-32-1)) = (1 \ll 31)$ or $2^{31}$ in binary-exponential notation. The for-loop in Algorithm~E now runs from bit $i_0-1= 4$ down to bit 0 of the remaining portion of ${\rm NOT}(p')$ (labeled $pshift$ in the pseudocode), yielding the Montgomery-mul, mod-add and binary-power sequence given in Table~7.
\begin{table}[ht]
\begin{center}
\label{pow_neg}
\caption{Sequence of mod-powers computed in inner loop of Algorithm~E, for $p = 977$, with $R=2^{64}$.}
\begin{tabular}{c|c|l|l|l}
	bit		&$i$th bit	&									&								&								\\
index $i$	&of NOT$(p')$	&\qquad function calls			&	\qquad operation			&	\qquad pow\_out				\\
\hline
	4		&0		&$\ \ \ \ \ {\rm MONT\_SQR}(2^{31})		$&$2^{  31\times2 - 64}			$&$2^{  -2}						$\\
	3		&1		&$2 \times	{\rm MONT\_SQR}(2^{-1})		$&$2^{  -2\times2 - 64}\times2	$&$2^{ -68}\times2 = 2^{ -67}	$\\
	2		&1		&$2 \times	{\rm MONT\_SQR}(2^{-66})	$&$2^{ -67\times2 - 64}\times2	$&$2^{-198}\times2 = 2^{-197}	$\\
	1		&1		&$2 \times	{\rm MONT\_SQR}(2^{-196})	$&$2^{-197\times2 - 64}\times2	$&$2^{-458}\times2 = 2^{-457}	$\\
	0		&0		&$\ \ \ \ \ {\rm MONT\_SQR}(2^{-457})	$&$2^{-457\times2 - 64}			$&$2^{-978}$
\end{tabular}
\end{center}
\end{table}

\noindent The unconditional mod-doubling of the powering-loop output yields $s = 2^{-977}$ (mod $q$) $= 7143819210136784550$. Taking the positive-power remainder $r = 2^{977}$ (mod $q$) $= 8623243291871090712$~-- which is just our previously computed remainder $(2^{977} - 1)\ {\rm mod}\ q$, plus 1~-- it is easily verified that $s \cdot r \equiv 1$ (mod $q$). This needs just 5 calls to MONT\_SQR; Note that this example also illustrates the worst-case scenario for the inverse-powering scheme, which is when the scaled power $p+b$ has one more bit than $p$ alone, leading to an extra loop execution relative to a more-conventional positive-powering ladder.

In order to make a more precise work comparison, let us compare this to an orthodox $2^p$\ (mod $q$) computation, also using the Montgomery modmul. Here we start with $p = 977$, which in binary form $= 1111010001$. Using $R = 2^{64}$ We again process the leftmost 6 bits of this {\em en bloc}; these $= 111101 = 61$. The required powering-loop-entry seed value is thus $s = R*2^{61}$ (mod $q$) $= 696971437655893565$. We can compute this in similar fashion as we did our seed value for the Algorithm~C scaling-factor computation, by first computing $R^{3/2}$ (mod $q$) via 64-bit floating-point arithmetic and calling MONT\_SQR to obtain $R^2$ (mod $q$), then calling MONT\_MUL($R^2$ (mod $q$), $2^{61}$) to obtain $s$. This is followed by 4 loop iterations, summarized in Table~8, and a final MMUL\_ONE of the loop output yields $2^{1041-64} = 2^{977}$ (mod $q$).Thus, while we have one fewer loop iteration in this case (and this is the exception rather than the rule) which saves us a mod-squaring, we incur three added modmuls (one squaring, one general multiply, one ``downshift multiply" by unity) in the pre- and postprocessing scalings. Thus in this example~-- which admittedly has a relatively small exponent $p$~-- the inverse-power scheme captured in Algorithm~E needs 5 modmuls (all squarings), compared to 7 total modmuls (5 of which are squarings) for the positive-power computation. The asymptotic cost for both schemes for large $p$ is of course the same: $O(\lg p)$ mod-squarings, and a similar number of mod-doublings, the latter of which are of negligible cost in the large-modulus-asymptotic sense.
\begin{table}[ht]
\begin{center}
\label{pow_pos}
\caption{Sequence of mod-powers computed in inner loop of standard Montgomery powering ladder to yield $2^p$ (mod $q$), for $p = 977$, with $R=2^{64}$.}
\begin{tabular}{c|c|l|l|l}
	bit		&$i$th bit	&							&								&								\\
index $i$	&of $p$		&\qquad function calls		&	\qquad operation			&	\qquad pow\_out				\\
\hline
	3		&0		&$\ \ \ \ \ {\rm MONT\_SQR}(2^{125})	$&$2^{125\times2 - 64}			$&$2^{ 186}						$\\
	2		&0		&$\ \ \ \ \ {\rm MONT\_SQR}(2^{186})	$&$2^{186\times2 - 64}			$&$2^{ 308}						$\\
	1		&0		&$\ \ \ \ \ {\rm MONT\_SQR}(2^{308})	$&$2^{308\times2 - 64}			$&$2^{ 552}						$\\
	0		&1		&$2 \times	{\rm MONT\_SQR}  (2^{552})	$&$2^{552\times2 - 64}\times2	$&$2^{1040}\times2 = 2^{1041}	$
\end{tabular}
\end{center}
\end{table}


If the additive constant $a \ne\pm 1$ we must compute the inverse $s^{-1}$ (mod $q$) via $b$-bit extended gcd in order to find the true remainder $2^p + a$ (mod $q$), which negates any savings resulting from the lack of explicit $R$-related scalings, but if $a = \pm 1$ and we only care about divisibility, we are done. As such, the method is ideally suited for finding or verifying factors of Fermat and Mersenne numbers.

The above inverse-powering algorithm, together with multiply arithmetic using $R = 2^{96}$, was used by the author to discover the fourth known (and currently largest-known) prime factor of the double-Mersenne number MM31, the 78-bit factor $q = 178021379228511215367151 = 2 \cdot 41448832329225 \cdot M_{31} + 1$. \cite{WMMM,MM31}


\section{Note Added in Proof}

Knuth~\cite{Knuth} hints at a right-to-left remaindering algorithm (Problem 4.3.1-41), but as this is cast as an exercise, there are scant details. Much more detail is provided in the paper by Eldridge \& Walter~\cite{EldWalt93}, who present both a right-to-left remaindering algorithm and a method for computing the quotient with it in ``on the fly'' fashion. Their quotient computation is not fully one-step deterministic in that it involves an initial estimate and a post hoc conditional error-correction step, but nonetheless allows for both remainder and quotient to be obtained with a single pass through the data. In that light, the main advantages of the present two-pass approach is its parallelizability~-- whether in ``overlapping instruction stream'' fashion as detailed here, or via explicit multithreaded implementation, or a combination of both. That parallelism is a direct result of the separate initial remaindering step, which is easily modified to yield as many ``partial remainders'' as desired, for either standalone recombination or for passing to a similarly parallel quotient-computation step. The author thanks an anonymous reviewer for the reference to the latter paper.


\section{Acknowledgements}

The fast divisibility Algorithm~A was first implemented by the author circa 1998 and used extensively since then in his experimental number theory codes, but was not considered by him to be suitable for publication without accompanying true-remainder and quotient algorithms. The latter came about only recently, thanks to an unexpected bout of free time sufficient to revisit the earlier work, for which the author would like to thank Synopsys Inc.

The author would like to thank Peter Montgomery, at whose metaphorical feet he first learned the basics of the latter's modmul method, and the ``hard code" crew at mersenneforum.com~-- George Woltman, Robert Holmes and Ross Schiff~-- for kindly taking the time to review an early version of this manuscript and providing numerous helpful x86\_64-coding-related insights.

\bibliographystyle{acmsmall}
\bibliography{mont_div_arxiv}

\begin{thebibliography}{}

\bibitem{Bar87}
{\sc Barrett, P.} 1987.
\newblock {Implementing the Rivest Shamir Adleman public key encryption algorithm on a standard digital signal processor}.
\newblock In {\em Advances in Cryptology.\ Proc.\ Crypto `86}, {A.~Odlyzko}, Ed. Lecture Notes in Computer Science Series, vol. 263. Springer, New York, 311--323.

\bibitem{CP05}
{\sc Crandall,R.} {\sc \&} {\sc Pomerance,C.} 2005.
\newblock {\em {Prime Numbers: A Computational Perspective}\/} 2nd Ed.
\newblock Springer.

\bibitem{EldWalt93}
{\sc Eldridge, S.} {\sc \&} {\sc Walter, C.} 1993.
\newblock {Hardware Implementation of Montgomery's Modular Multiplication Algorithm}.
\newblock {\em IEEE Trans.\ Comp.\/}~{\em 42[6]}, \mbox{693--699}.

\bibitem{agner}
{\sc Fog, A.} 2013.
\newblock {Instruction tables: Lists of instruction latencies, throughputs and micro-operation breakdowns for Intel, AMD and VIA CPUs}.
\newblock {\em http://www.agner.org/optimize}.

\bibitem{GMP}
\newblock {GNU Multiple Precision Arithmetic Library manual}.
\newblock {\em http://gmplib.org/gmp-man-5.0.5.pdf}.

\bibitem{GMPtiming}
\newblock {GMP instruction timing assembly chart}.
\newblock {\em https://gmplib.org/devel/asm.html}.

\bibitem{Hensel}
{\sc Hensel, K.} 1908.
\newblock {\em {Theorie der algebraischen Zahlen}}.
\newblock Teubner, Leipzig.

\bibitem{Knuth}
{\sc Knuth, D.~E.} 1981.
\newblock {\em {Seminumerical Algorithms}\/} 2nd Ed.\ (TAOCP Series, vol.~2).
\newblock Addison-Wesley.

\bibitem{WMMM}
{\sc MathWorld,}
\newblock {Double Mersenne Number}.
\newblock {\em http://mathworld.wolfram.com/DoubleMersenneNumber.html}

\bibitem{MM31}
{\sc Mayer, E.} 2005.
\newblock {Fourth Known Factor of M(M31). NMBRTHRY mailing list}.

\bibitem{Mene96}
{\sc Menezes, A.}, {\sc van Oorschot, P.}, {\sc \&} {\sc Vanstone, S.} 1996.
\newblock {\em {Handbook of Applied Cryptography}}.
\newblock CRC Press, Inc., Boca Raton, FL, USA.

\bibitem{GMP_div}
{\sc M\"oller, N.} {\sc \&} {\sc Granlund, T.} 2010.
\newblock {Improved division by invariant integers}.
\newblock {\em gmplib.org/$\sim$tege}.

\bibitem{Mont85}
{\sc Montgomery, P.} 1985.
\newblock {Modular multiplication without trial division}.
\newblock {\em Math.\ Comp.\/}~{\em 44}, \mbox{519--521}.

\end{thebibliography}


\end{document}